\documentstyle[pra,epsfig,floats,aps]{revtex}
\begin{document}

\title{Theoretical study of collective modes in DNA at ambient temperature}
\author{Simona Cocco $^{1,2,3}$ and R{\'e}mi Monasson
$^3$ 
}
\address {$^1$  Dipartimento di  Scienze Biochimiche,
Universit{\`a} di Roma "La Sapienza",
P.le A. Moro, 5 - 00185 Roma, Italy}
\address{$^2$ CNRS-Laboratoire de Physique de l'ENS-Lyon,
46 All{\'e}e d'Italie, 69364 Lyon Cedex 07, France.}
\address{$^3$ CNRS-Laboratoire de Physique Th{\'e}orique de l'ENS,
24 rue Lhomond, 75005 Paris, France.}   
\maketitle
\begin{abstract}
The instantaneous normal modes corresponding to base pair vibrations
(radial modes) and twist angle fluctuations (angular modes) of a DNA
molecule model at ambient temperature are theoretically investigated.
Due to thermal disorder, normal modes are not plane waves with a
single wave number $q$ but have a finite and frequency dependent
damping width. The density of modes $\rho (\nu )$, the average
dispersion relation $\nu (q)$ as well as the coherence length $\xi
(\nu)$ are analytically calculated. The Gibbs averaged resolvent is
computed using a replicated transfer matrix formalism and variational
wave functions for the ground and first excited state. Our results for
the density of modes are compared to Raman spectroscopy measurements
of the collective modes for DNA in solution and show a good agreement
with experimental data in the low frequency regime $\nu <
150$~cm$^{-1}$. Radial modes extend over frequencies ranging from
50~cm$^{-1}$ to 110~cm$^{-1}$.  Angular modes, related to helical axis
vibrations are limited to $\nu < 25$~cm$^{-1}$. Normal modes are
highly disordered and coherent over a few base pairs only ($\xi < 20
{\AA}$) in good agreement with neutron scattering experiments.
\end{abstract}

%
%

\section{Introduction}

Much attention has been paid in the last twenty years to low frequency
dynamics of DNA due to its relevance to biological processes.
Vibrational modes involving collective motions of groups of atoms have
been experimentally investigated by means of various techniques as
neutron scattering~\cite{Gri87,Gri91,Gri97}, spectroscopy
measurements~\cite{Pai81,Ura81,Ura83,Ura85,Lin88}, NMR,~... Among the
latters, Raman studies carried out by H.~Urabe~{\em et
al.}\cite{Ura81,Ura83,Ura85,Ura91} have revealed particulary useful to gain
information on the dependence of low frequency vibrations properties
of DNA upon external conditions e.g. water content, ionic
concentration, temperature. The observation of the Raman scattering
intensity in the low frequency region is technically very difficult
because the strong solvent Rayleigh scattering near zero frequency
masks the DNA response.  Nevertheless a broad band ranging from 
60 cm$^{-1}$ to $100$~cm$^{-1}$ has been evidenced and associated to hydrogen
bonded base pair vibrations~\cite{Ura81,Ura83}. Moreover experiments
focusing on oriented solid DNA fibers have also exhibit sharp peaks in
the Raman intensity at $\simeq$~16~cm$^{-1}$, that shifts toward
lower frequency region when rising the degree of
hydration~\cite{Ura85}. The origin of this peak is far from being
obvious. In particular, its interhelical or intrahelical origine has
been debated for a long time~\cite{Ura91}.

To reach a better understanding of the above experimental findings,
detailed theoretical analysis of DNA vibrational motions have been
proposed. Devoting a particular attention to hydrogen bond stretching
modes, Prohofsky and his collaborators~\cite{Pro95} have been able to
find back a vibration mode at 85~cm$^{-1}$ and confirm its origin. Their
approach is based on a detailed description of the DNA molecule at the
atomic level~\cite{Eys74} and a variational calculation method, called
modified self-consistent phonon approximation (MSPA). Within MSPA, the
molecule at ambient temperature is reduced to an effective (and
complex) harmonic lattice the force constants of which are determined
in a self-consistent way. This approach has been very useful in
calculating many properties of DNA, e.g. the melting of DNA's of
different sequences \cite{Gao84},\cite{Che94}, the temperature and
salt dependence of the B to Z conformation change \cite{Che94b} and
the stability of drug-DNA complex \cite{Che97}.

A qualitative weakness of MSPA, that also arises in other simplified
theoretical approaches~\cite{Lis96} lies in the {\em a priori} nature
of modes.  Though effective elastic constants depends on temperature,
normal modes indeed conserve a plane wave structure~\cite{Pro95}.  As
a consequence, dispersion relations for the normal modes are well
defined as for phonons in crystals~\cite{Kit83}. In other words, the
coherence length is infinite and the momentum selection rule give rise
to a discrete set of lines in the theoretical predictions for Raman
spectra\cite{Kit83}. On the contrary, amorphous materials or more
generally thermally disordered systems give rise to normal modes with
short coherence lengths, whose power spectrum displays a finite
width~\cite{Gri91,Gri97,Gri88}.  As a result, momentum selection rules
break down and light scattering processes may occur from essentially
all normal modes~\cite{Shu70,Mar74}. Correspondingly for a DNA
molecule in solution, the Raman spectrum is continuous and does not
depend on the wave length of the incident laser beam nor on the
scattering angle~\cite{Ura81}.
 
Understanding the vibrations of a system in presence of thermal (or
configurational) disorder is a difficult task~\cite{Sou86,Ell90}.
Non-harmonicities in the interactions between constituents may modify
deeply the usual phonon picture of wave propagation in solids and
generally forbid any rigorous analytical treatment of the dynamical
equations. An interesting approach to circumvent this difficulty has
been proposed in the context of liquid state
dynamics~\cite{Key97}. The idea is to start from a randomly chosen
configuration at thermodynamical equilibrium. Then, the equations of
motion around this fixed initial configuration can be linearized,
defining some instantaneous normal modes (INM) and their corresponding
relaxation times {\em i.e.} frequencies.  All these quantities may
then be averaged over the initial configuration chosen from the Gibbs
ensemble at the desired temperature.  One obtains this way a precise
description of the short times dynamics based on the average spectrum
of relaxation times and the statistical properties of the
INM~\cite{Key97}. As a major advantage, the INM approach takes into
account thermal disorder through the choice of the random initial
configuration and gives rise to highly disordered normal modes with
finite autocorrelation length as expected at finite temperature.  Due
to the linearization procedure, the information available through INM
calculations is restricted to short times dynamics. This limitation is
however not serious in the range of frequencies mentioned above for
DNA collective dynamics as we shall see later.
   
From a theoretical point of view, the calculation of the INM around a
fixed intial configuration is a hard task that can be performed
analytically for simple enough models only\cite{Wan94,Bir99,Fig99}. 
In this paper we describe
the torsional and hydrogen bond stretching vibrations of a very simple
DNA model by normal modes that are not plane waves using the INM
approach.  We are able in particular to calculate the frequency
dependent density of state at a given temperature, the coherence
length of the normal modes and some pseudo dispersion relations at
ambient temperature.  The model that we consider has been introduced
in a previous article to study the DNA denaturation driven by
temperature or mechanically induced by a torque~\cite{cocco}. This
study was motivated by the recent development of micromanipulation
techniques\cite{Smi92,Smit92} that have allowed the direct observation of DNA
denaturation induced by an external torsional stress~\cite{Str96,Stri99}.  Our
model describes the DNA molecule at the base pair level, and for each
base pair we have just two degrees of freedom that is the base pair
radius and twist angle. As a result, the potential energy is
sufficiently simple to allow for sophisticated analytical calculations
of normal collective modes.  From a technical point of view, our
calculation is inspired from a recent study of instantaneous modes in
an one dimensional disordered system~\cite{Mon99}, that mixes
techniques used in localization theory (the resolvent calculation),
disordered system (replica trick)\cite{Tho95} and a variational Gaussian wave
function method. As we shall see, the interest of the calculation is
two-fold. First, it gives theoretical predictions for the spectrum of
modes, the effective dispersion relations and damping width at ambient
temperature that can be directly confronted with Raman spectroscopy
and to neutron scattering data. Secondly, it is a way to check the
validity of (and eventually improve) our model in very different
experimental conditions from the the ones it was originally designed
for.

The paper is organized as follows. In Section~II, we define the DNA model
and explain how to compute its statistical physics properties. The
choice of the force constants and the main thermodynamical features
are then exposed. Section~III is devoted to dynamics and to the
definiton of the instantaneous modes. The relationship between the
latters and Raman spectra is evoked. The analytical framework
necessary to calculation the properties of INM is presented in
Section~IV. Results are given and compared to experiments in
Section~V. Finally, we present some conclusions and perspectives in
Section~VI.

%
%

\section{Presentation and main features of the DNA model}

In this Section, we describe the DNA model under study and 
briefly recall how it can be studied using statistical mechanics techniques
as well as its main thermodynamical features, see \cite{cocco}. 
We then discuss the properties of the
normal modes, e.g. dispersion relations, spectral density, ...
of the molecule in the absence of temperature and interaction
with the solvent.

\subsection{Definition of the model}

Our model reproduces the Watson-Crick double helix (B-DNA) as
schematized fig~1. Each base pair ($j=1,\ldots ,N$) is described
by its radius $r_j$ and the angle $\varphi_j$ 
in the plane perpendicular to the helical axis\cite{Bar99}. 
The sugar phosphate backbone is made of rigid
rods, the distance between adjacent bases on the same strand being
fixed to $L =6.95 {\AA}$. Conversely, the distance $h_j$ between 
base planes $j-1$ and $j$ is allowed to fluctuate.
The Hamiltonian $V$  associated to a configuration of the
degrees  of freedom $\{ r_j ,
\varphi _j \}$ is the sum of three different contributions, 
\begin{equation}
V [\{ r_j , \varphi _j \} ] = \sum _{j=1} ^N V_m (r_j) +
\sum _{j=2} ^N \left( V_s ( r_j , r_{j-1} ) + 
V_b (r _j , r _{j-1} , \varphi _j - \varphi _{j-1} ) \right) 
\qquad , \label{hamil}
\end{equation}
that we now describe.

\begin{itemize}
\item Hydrogen bonds inside a given pair $n$ are taken into
account through the short-range Morse potential \cite{Pro95,PB89} 
\begin{equation}
V_m (r_j)=D\, \left(e^{-a(r_j-R)}-1\right)^2
\end{equation}
with  $R=10{\AA}$ .
The width of the well amounts to $3a^{-1} \simeq 0.5 {\AA}$ 
\cite{Daux95}, in agreement with the 
order of magnitude of the relative motion of the hydrogen bonded bases 
\cite{Mac87}. A base pair with diameter $r > r_d = R + 6/a$ may be
considered as open. The potential depth $D$, typically of the order of 
$0.1 eV$ \cite{Pro95} depends on the base pair type 
(Adenine-Thymine (AT) or Guanine-Citosine (GC)) as well as on the 
ionic strength. Note that the Morse potential $V_m$ increases 
exponentially with decreasing $r< R$ and may be considered as infinite
for $r< r_{min} = 9.7 {\AA}$ \cite{Zhan97}.

\item The shear force that opposes sliding motion of one base over
another in the B-DNA conformation is accounted for by the stacking
potential \cite{Saen84} 
\begin{equation}
V_s(r_j,r_{j-1})= E\, e^{-b (r_j +
r_{j-1}-2R)}\; (r_j-r_{j-1})^2 \qquad .  
\end{equation}
Due to the decrease of molecular
packing with base pair opening, the shear prefactor is exponentially
attenuated and becomes negligible beyond a distance $\simeq 5 b^{-1} =
10 {\AA}$, which coincides with the diameter of a base pair
\cite{Daux95}.

\item An elastic energy is introduced to describe the vibrations of 
the molecule in the B phase, 
\begin{eqnarray}
V_b (r _j , r_{j-1} , \theta _j ) &=& K \big( h_j - H \big) ^2
\nonumber \\ &=&  K \left(  \sqrt{ L^2 - r_j ^2 -
r_{j-1}^2 +2 r_j r_{j-1} \cos \theta _j } - H \right)^2 
\label{vb}
\end{eqnarray}
where $\theta _j = \varphi _j - \varphi _{j-1}$ is the twist angle
between base pairs $j-1$ and $j$.
The helicoidal structure arises from the choice of $H < L$: in
the rest configuration $r_j=R$ at $T=0$K, $V_b$ is minimum and zero
for the twist angle $\Theta > 0$ with $\sin (\Theta/2) =
\sqrt{L^2-H^2}/(2R)$. The above definition of $V_b$ holds as long as
the argument of the square root in (\ref{vb}) is positive, that is if
$r_j , r_{j-1} , \theta _j$ are compatible with rigid rods having
length $L$. By imposing $V_b = \infty$ for negative arguments,
unphysical values of $r _j , r_{j-1} , \theta _j$ are excluded. As the
behaviour of a single strand ($r > r_d$) is uniquely governed by this
rigid rod condition, the model does not only describe vibrations of
helicoidal B-DNA but is also appropriate for the description of the
denaturated phase \cite{cocco}.
\end{itemize}

\subsection{Calculation of partition function}

The configurational partition function at inverse temperature $\beta =
1/ (k_B T)$ reads
\begin{equation}
\label{ZG}
Z = \int _{r_{min}} ^\infty r_1 dr_1 \int _{-\infty} ^\infty d\varphi _1 
\ldots \int _{r_{min}} ^\infty r_N dr_N \int _{-\infty} ^\infty d\varphi _N 
\  \exp \bigg\{ -\beta V[ \{ r_j , \varphi _j \}] \bigg\}\; 
\delta (\varphi _1) \prod _{j=2}^{N} 
\chi \big( \theta _j  \big) \ .
\end{equation}
The angle of the first extremity 
of the molecule is set to $\varphi_1=0$ with no restriction (due to
the arbitrary choice of the angular reference axis, see fig~1) whereas
the last one $\varphi _N$ is not constrained. 
The $\chi$ factors entering (\ref{ZG}) are defined by
$\chi(\theta _j ) =1$ if $0\leq \theta _j=
\varphi _j-\varphi _{j-1} \leq \pi$ and 0 otherwise to prevent any clockwise
twist of the chain.
Partition function $Z$ can be calculated using the transfer integral
method\cite{parisi},
\begin{equation} \label{ZGP}
Z = \int _{r_{min}} ^\infty r_1 dr_1 \int _{r_{min}} ^\infty r_N dr_N 
\int _{-\infty} ^\infty d\varphi _N \ 
\langle r_N , \varphi _N | T^N| r_1 , 0 \rangle \quad ,
\end{equation} 
where the transfer operator entries read $\langle
r,\varphi|T|r',\varphi' \rangle = T(r,r',\theta )$ with 
$\theta = \varphi - \varphi '$ and
\begin{eqnarray}
T(r,r',\theta ) &=& X(r,r')\, \exp \left\{
-\beta V_b \,(r,r', \theta ) \right\} \, \chi \big( \theta \big) \quad ,\\
X(r,r') &=& \sqrt{rr'}\; \exp \left\{ -\frac{\beta}{2} \left( 
 V_m\,(r)+  V_m\,(r') \right) -\beta  V_s\,(r,r') \right\} \qquad . 
\label{deft}
\end{eqnarray} 
At fixed $r,r'$, the angular part of the transfer
matrix $T$ is translationally invariant in the angle variables
$\varphi$, $\varphi '$ and can be diagonalized through a Fourier
transform. Thus, for each Fourier mode $k$ we are left with an
effective transfer matrix on the radius variables
\begin{equation} \label{autremu}
T_k(r,r')=X(r,r')\,Y_k(r,r')
\end{equation}
with
\begin{equation}
\label{mu}
Y_k(r,r') = 
\int_{0}^{\pi}\,d\theta \exp \bigg\{ -\beta  V_b\,(r,r',\theta) 
-i k \theta \bigg\} \quad .
\end{equation} 
The only mode contributing to $Z$ is $k=0$ once $\varphi _N$
has been integrated out in (\ref{ZG}).  In the $N\to\infty$
limit, the free-energy density $f$ is simply given by $f
= -k_B T \ln \lambda _{0}$, where $\lambda _{0}$ is
the maximal eigenvalue of $T_0$ whose corresponding eigenvector will
be denoted by $\psi _{0}(r)$.

\subsection{Thermodynamical properties and parameters}

The ground state wave function $\psi _0 (r)$ is shown figure~2
at ambient temperature $T=300$~K. It is entirely localized in the
Morse potential well as expected for a DNA molecule in B
configuration. The first excited eigenstate of the transfer matrix
has an excess free-energy $\Delta G$ with respect to $\psi _0$ and is 
delocalized: it extends over all values of $r > r_d$, vanishes for
$r<r_d$ and thus represents a denaturated molecule. 
At some higher temperature $T_m$, the bound wave function disappears
and $\psi _0$ suddenly undergoes a delocalization transition. In
other words, hydrogen bonds break up and $T_m$ can be interpreted 
as the melting temperature\cite{cocco}.

The values of the parameters entering the potential energy is
discussed in \cite{cocco} and  are listed below:
\begin{itemize}
\item inverse hydrogen bond length: $a=6.3 {\AA}^{-1}$.
\item zero temperature interplane distance: $H=3{\AA}$.
\item Morse potential depth: $D=0.16 eV$.
\item attenuation coefficient for stacking interactions: $b=0.49 {\AA}^{-1}$.
\item stacking stiffness: $E=4 eV/{\AA}^2$.
\item backbone elasticity constant: $K=0.014 eV/{\AA}^2$. 
\end{itemize}
The values of $a$ and $b$ have been borrowed from literature
\cite{cocco}.  The choice of the other parameters $D,E,K,H$ ensures
that geometrical and thermodynamical properties as the average twist
angle, the mean axial distance between successive bases in the B
conformation, the melting temperature $T_m=350$~K and the denaturation
free-energy $\Delta G$ are correctly predicted\cite{cocco}.

Notice that the main uncertainty in this tuning procedure arise with
the choice of the stacking stiffness $E$. Three
possible pairs of parameters $(D,E)$ that correctly fit $T_m =350$~K are
listed in Table~\ref{tavb}, as well as the corresponding denaturation
free-energies $\Delta G$ at $T=300$~K. We have selected the pair
 giving the largest prediction for the
denaturation free-energy that is in closest agreement with
thermodynamical estimates of $\Delta G$\cite{Saen84}. 
It can be easily seen from
Table~\ref{tavb} that $E$ varies much more than $D$ and $\Delta G$ and
is therefore less accurately predicted than the other parameters of
the model. 

%
%

\section{Dynamics and Instantaneous Normal Modes}

Our aim is to perform an analytical calculation of the spectrum of
torsional and radial vibrations on characteristic time scale of the
pico seconds and at ambient temperature.  In this section, we first
write the dynamical equations for the model. We then linearize these
equations around a given (and randomly chosen) configuration of the
thermally equilibrated system and define the instantaneous normal
modes as the vibrations of the DNA molecule around this
configuration. The density of instantaneous normal modes can be related
to some extent to the Raman intensity.  Finally, we consider the
special case of zero temperature. The explicit calculation of the
dispersion relations and the density of modes allows for a decoupling
of the angular and radial motions. It will also provides useful
comparisons with the finite temperature results of Section V.

\subsection{Dynamical equations for the DNA model}

We denote the configuration of the molecule at time $t$ by $\{ r^{t}_{j} , 
\varphi^{t}_j \}$. The equations of motion read
\begin{eqnarray}
\label{eqm}
m \, {\ddot r}^{t}_{j} &=&  m\, r^{t}_{j} ({\dot \varphi}^{t}_{j})^{2}
 -\frac{1}{2} 
\frac{\partial V}{\partial r^{t}_j} \nonumber  \\ 
m\,(r^{t}_j)^2 {\ddot \varphi}^{t}_j&=&-\frac{1}{2}  \frac{\partial
 V}{\partial \varphi^{t}_j} \quad , \label{motion}
\end{eqnarray}
where the potential energy $V$ has been defined in (\ref{hamil}). 
The effective half mass $m = 300$~u.m.a. of a base pair takes into account the
atomic consituents of the nucleotide and of the backbone as well as 
the primary hydration shell which is 
tightly bound to the base~\cite{Lis96,Saen84,Ura98}. The size
of the primary shell depends on the hydration degree and is
of the order of 10-20 water molecules per nucleotide. The
characteristic relaxation time of the primary shell is typically 
$\tau_1 \simeq 10^{-10}$~s \cite{Ura98}. Therefore, for dynamical
processes taking place on time scales $\tau < \tau_1$, that is for
frequencies $\nu > 0.3$~cm$^{-1}$  primary shells may be considered
as rigidly linked to the bases and simply taken into account
through the effective mass $m$. Our choice for $m$ is in close agreement
with the mass considered by Volkov and Kosevich~\cite{Vol91}. Taking
into the additional water shell mass and averaging over the possible
bases A,T,G and C, these authors have calculated somes estimates of the
masses of the nucleotide $m_n=199$~u.m.a. and of the backbone elements
$m_b=109$~u.m.a. giving a total mass $m=m_n+m_b=308$~u.m.a. per
base~\cite{Vol91}. 

The secondary hydration shell contains less rigidly bound water
molecules that induce some friction acting on the DNA molecule.  The
characteristics scale time of viscosity dissipation at ambient
temperature is $\tau_2 \simeq
10^{-12}$~s~\cite{Ura98,Tom85}, that is of the same order of magnitude
as in bulk water. The viscous terms $-\gamma \;
\dot{r}_j^t$ and $-\gamma \; \dot{\varphi}_j^t$ that should be included in
the motion equations (\ref{motion}) can be neglected on time scales 
$\tau< \tau_2 $ that is for frequencies $\nu>30$~cm$^{-1}$ only.
In the following we shall use the non dissipative equations
(\ref{motion}) keeping in mind that the interpretation of the results
must be made with care for frequencies smaller than 30~cm$^{-1}$. 

\subsection{Linearization approximation and instantaneous modes}

We linearize the equations of motions (\ref{eqm}) around a casual initial
configuration ${\cal C} = \{ r_j,\varphi_j\}$ of the system already in
thermodynamical equilibrium, defining for each base pair $j$, 
\begin{eqnarray}
 r^{t}_j&=& r_j+y^{t}_j \nonumber \\
\varphi^t_j&=& \varphi_j+\tilde{\phi}^t_j.
\end{eqnarray}
Once linearized the dynamical equations (\ref{eqm}) 
are rewritten in terms of the displacement
variables $ y^{t}_j , \phi_j ^t \equiv R . \tilde{\phi^t_j}$ and read
\begin{eqnarray} 
\label{lin}
m \, {\ddot y}_{j} ^t &=& \sum_k ({\cal D}^{\cal C}_r)_{j,k}\,y_k ^t + \sum_k
({\cal D}^{\cal C} _m)_{j,k}\,\phi_k ^t \nonumber \\
m \, {\ddot \phi}_{j} ^t &=& \sum_k ({\cal D}^{\cal C}
_{m})_{k,j}\,y_k ^t + \sum_k
({\cal D}^{\cal C} _{\varphi})_{j,k}\,\phi_k ^t 
\end{eqnarray} 
where 
\begin{equation}
 \big( {\cal D}^{\cal C} _r  \big) _{jk}
= \left.\frac 12 \frac{\partial^2 V}{\partial r_j \partial
r_k }\right|_{\cal C} \ , \qquad
\big( {\cal D}^c_m \big) _{jk}= \left. \frac{1}{2R} \frac{\partial^2 V}
{\partial r_j \partial \varphi _k}\right|_{\cal C}  \quad
\hbox{\rm and} \qquad
\big(
{\cal D}^c_{\varphi}\big) _{jk} = \left. \ \frac{1}{2R^2}\frac{\partial^2 V}{
 \partial \varphi_j \partial \varphi _k} \right|_{\cal C}
\end{equation}
are the $N\times N$  matrices of second derivatives of the potential
energy $V$ around configuration ${\cal C}$.
To solve the linear system (\ref{lin}) one has to find the  eigenvalues
$\lambda$ of the $2N \times 2N$ Hessian matrix 
\begin{eqnarray}
\label{Hes}
{\cal D}^{\cal C}= 
\left( \begin{array}{cc} 
{\cal D}^{\cal C}_r & {\cal D}^{\cal C}_m\\
{\cal D}^{\cal C}_m & {\cal D}^{\cal C}_{\varphi}
\end{array} \right) \qquad .
\end{eqnarray}
 It is important to notice that due to the dependence on the initial
configuration ${\cal C}$ the elements of the matrix ${\cal
D}^{\cal C}$ are not translationally invariant.  Consequently, the
eigenvectors of ${\cal D}^{\cal C}$, {\em i.e.} the instantaneous
normal modes corresponding to
the initial configuration ${\cal C}$ are not plane waves.

The eigenvalues histogram $\rho^{\cal C} (\lambda)$ give the density of the
normal modes as a function of the initial configuration. The latter is
distributed according to the Gibbs measure
\begin{equation} \label{gibbs}
{\cal P} ({\cal C} = \{r_j ,\varphi _j \} ) 
= \frac 1{Z} \; \exp \left( -\beta 
V [\{ r_j ,\varphi _j  \} ] \right) ,
\end{equation}
where $V$ is the Hamiltonian defined in equation (\ref{hamil}).
Once averaged over distribution (\ref{gibbs}), the mean 
density of states $\rho (\lambda)$ is available and depends only on
the temperature $T=1/\beta$. The frequency spectrum $\rho _f (\nu)$ 
is straightforwardly obtained through the relationship, see
(\ref{lin}),
\begin{eqnarray}
\label{ris}
\nu &=& \frac 1{2\pi}\;\sqrt{ \frac {\lambda }{m}} \nonumber \\
\rho _f (\nu )  &=& 4 \pi \sqrt {m \; \lambda } \; \rho (\lambda )
\qquad . \label{eq}
\end{eqnarray}
To lighten notation, we shall drop the $f$ subscript in (\ref{eq}) and
use indifferently the same letter $\rho$ to denote the density of
states as a function of the eigenvalue $\lambda$ or frequency $\nu$.

The spectrum of the Hessian matrix (\ref{Hes}) is not necessarily
positive. Some instantaneous modes may be unstable and grow with time
within the linear approximation. Their corresponding eigenvalues
$\lambda$ are negative and the associated frequencies $\nu$ are purely
imaginary. Following the conventions of \cite{Key97}, imaginary
frequencies $\nu$ will be conveniently represented by minus their 
modulus $-|\nu |$, that is by points on the negative frequency semi-axis. 
We shall come back in Section~V to these modes.

\subsection{Relationship with Raman spectra}

In disordered solids the Raman scattering intensity is directly
related to the density of normal modes.  Indeed when the coherence
length of the normal modes is short compared to optical wavelengths
the conservation of momentum is no longer a restrictive selection rule
and does not give rise to a discrete set of lines for the spectrum.
Light scattering processes occur from essentially all the normal modes
of the material and the spectra are continuous.  As we will see later,
the assumption of short coherence length is well verified in our model
at finite temperature.  The relationship between the Raman intensity
${\cal I} (\nu )$ and the density of normal modes $\rho(\nu)$ is generally
expressed via the light-to-vibrations coupling coefficient $C(\nu)$
\cite{Shu70},
\begin{equation}
\label{rescaraman}
{\cal I}(\nu)= \rho (\nu)\; C (\nu) \; \left( \frac{ n(\nu)+1 }{ \nu}
\right)  \qquad ,
\end{equation}
where
\begin{equation}
n(\nu) = \frac 1{\exp \left( \frac{h \nu}{k_B T} \right) -1} 
\end{equation}
is the average population of level $\nu$.  The spectral dependence of
$C(\nu)$ is still an unsettled question.  Early studies devoted to the
relationship between Raman spectra and densities of modes assumed that
the light-to-vibrations coupling was indepent of frequency, {\em i.e.}
$C(\nu)= 1$\cite{Shu70}.  Later works conjectured a quadratic
behaviour $C(\nu)\sim \nu^2$ at very low frequencies and a less steep
increase for larger $\nu$\cite{Mar74}.
Recently, comparisons between neutron scattering experiments and
calorimetric measures have given evidence for a linear dependence
$C(\nu ) \propto \nu$ for different glasses in the frequency range 
$8\, cm^{-1}< \nu < 100\, cm^{-1}$\cite{Mal90,Sok93}, the upper bound
being related to the Debye frequencies of the corresponding crystals.

\subsection{Normal modes at zero temperature}

The linearized equations af motions at zero temperature are obtained
performing an expansion of the potential energy $V$ up to the second
order atound the rest positions: $r^t_j=R+y^t_j,\;
\varphi^t_j=n\Theta+ \phi^t_j/R$:
\begin{eqnarray}
\label{eqm1}
m\,{\ddot y^t}_j 
&=&-a^{2} D y^t_j 
-K_{yy}\, (2{y^t}_{j} + {y^t}_{j+1} + {y^t}_{j-1})-
E\,(2 y^t_j-y^t_{j+1}-y^t_{j-1}) - K_{y\,\phi}\,(\phi^t_{j+1}-\phi^t_{j-1})
\nonumber \\
\label{eqm2}
m{\ddot{\phi}}^t_{j} 
&=& - K_{\phi \phi}\, (2{\phi}^t_{j}-{\phi}^t_{j+1}-{\phi}^t_{j-1})+
K_{y\,\phi}\,(y^t_{j+1}-y^t_{j-1})
\nonumber \\
\end{eqnarray}
where
\begin{eqnarray}
K_{yy} &=&  \big( K R_0^2 / H^2 \big){(1 - \cos{\Theta})}^{2}\;,\\
K_{\phi\phi} &=&  \big( K R_0^2 / H^2 \big){(\sin^2{\Theta}_{0})}\;, \\
K_{y\phi}&=&\big(K R_0^2/ H^2 \big)(\sin{\Theta}_{0})(1-\cos{\Theta}_{0})\; 
\end{eqnarray}
The  plane waves
\begin{eqnarray}
\left( \begin{array}{c} y_j \\  \phi_j
 \end{array} \right)=\left( \begin{array}{c} y_\pm (q) \\  \phi_\pm (q)
 \end{array} \right) \exp \big\{ i ( q n-2\pi\nu_\pm (q) t )\big\}
\end{eqnarray}
are solutions of (\ref{eqm1}) with the relations of dispersion $\nu
_\pm (q)$ showed in fig~3.  Due to the difference of
the order of magnitude between $a^2 D =6.33\; eV\, {\AA}^{-2}$, $E=4\;
eV\, {\AA}^{-2}$ (or for the other choice of the parameters $a^2 D
=5.93\; eV\,{\AA}^{-2}$, $E=0.74\; eV\, {\AA}^{-2}$) and $K_{y
\phi}=18\;10^{-3}\;eV\, {\AA}^{-2}$, $K_{yy}=6\;10^{-3}\;eV\,
{\AA}^{-2}$, $K_{\phi\phi}=54\;10^{-3}\;eV\, {\AA}^{-2}$, it is clear that
the angular and radial motions take place on two different time scales
and become independent. The dispersion relations obtained when setting
$K_{y \phi}=0$ and $K_{y y}=0$ are indistinguishable from the previous ones
and read
\begin{eqnarray}
\nu _{r} (q) \equiv \nu_+ &=&\frac 1{2\pi} \sqrt{\frac
 {a^2D}{m}+2\,\frac{E}{m} (1-\cos{q})}\\
 \nu _{\varphi} (q) \equiv \nu_{-} &=&\frac 1{2\pi} \sqrt{\frac{2 
K_{\phi\phi}}{m} (1-\cos{q})} \label{disp67}
\end{eqnarray}
The density of states  $\rho (\nu)= 1 /(2\pi |\nu '(q)| )$ for each
branch is given by
\begin{eqnarray}
\rho _r (\nu )&=&\frac{4\pi \nu}{\sqrt{\left[ (2\pi\nu)^2 - \frac
 {a^2D}{m} \right] \left[ \frac{a^2D + 4 E}{m} - (2\pi\nu)^2 
\right]} } \\
\rho _\varphi (\nu )&=& \frac 1 {\sqrt{\frac {K_{\phi\phi}}{m}
- ( \pi \nu )^2}}
\quad . \label{density67}
\end{eqnarray}
Note that the above densities are both normalized to $1/2$ so that
the total density of states $ \rho _r (\nu ) +
\rho _\varphi (\nu )$ is properly normalized to unity.
As shown fig~4, Van Hove singularities are located at
frequencies $\nu$ such that the denominators in (\ref{density67}) vanish,
that is stationary points of the dispersion relations (\ref{disp67}).

\subsection{Structure of the Hessian matrix}

From the above considerations on the effective decoupling 
of radial and torsional modes, ${\cal D}^{\cal C}_m$ will be set
to zero hereafter. Therefore, the Hessian matrix (\ref{Hes}) is comprised of 
two $N\times N$ matrices, that is the radial Hessian matrix 
${\cal D}^{\cal C}_r$ and the angular Hessian matrix  
${\cal D}^{\cal C}_{\varphi}$. In what follows, the notation 
${\cal D}^{\cal C}$ will generically refer to either of these two matrices. 

Because only nearest neighbors along the molecule interact with each other, 
the Hessian matrix has a band diagonal structure,
\begin{eqnarray}
({\cal D}^{\cal C})_{kj} =
\left\{ \begin{array}{c c c}\frac 12 d_0 (r _j,r _{j-1}, \theta _j) +
\frac 12 d_0 (r_j, r_{j+1}, - \theta _{j+1})  & \hbox{\rm if } & k=j \\
-\frac 12 d_1 (r _j,r_{j-1},\theta _{j} ) & \hbox{\rm if } & k=j- 1 \\
- \frac 12 d_1 (r _{j+1},r_{j},\theta _{j+1} ) & \hbox{\rm if } & k=j + 1 \\
0 & \hbox{\rm if } & |k-j| \ge 2  \end{array} \qquad .
\right . \label{Hess}
\end{eqnarray}
Explicit expressions of elements $d_0$ and $d_1$ are:
\begin{itemize}
\item for the radius Hessian matrix ${\cal D}^{\cal C}_{r}$, the elements
\begin{eqnarray}
d_0 (r,r' ) &=& \frac 12
\frac{d^2 V_m}{dr^2} (r) +  2 E \qquad , \nonumber \\
d_1 (r,r' ) &=& 2 E \qquad ,
\label{d0d1radius} 
\end{eqnarray}
do not depend on the twist $\theta$.
For simplicity, we have not considered the exponential attenuation term
in $V_s$ which is almost constant (and equal to unity) in the B-DNA
phase. $K_{yy}$ has been set to zero since it is three orders of magnitude
smaller than other radial force constants.
\item for  the twist Hessian matrix
${\cal D}^{\cal C}_{\varphi}$, 
\begin{equation}
d_0 (r,r',\theta ) = d_1 (r,r',\theta )=
\frac 1{R^2} \;\frac{d^2 V_b}{d\theta ^2} (r,r',\theta ) \qquad ,
\label{d0d1twist} 
\end{equation}
where the backbone potential has been defined in (\ref{vb}).
\end{itemize}

%
%

\section{Analytical framework}

\subsection{Definition of spectral quantities}

We call $\lambda  ^{\cal C}_e$ (respectively $w ^{\cal C}
_{a,e}$) the eigenvalues
(respectively the components of the associated eigenvectors normalized
to unity) of ${\cal D} ^{\cal C} $, with $e=1,\ldots ,N$.
Most spectral properties of ${\cal D} ^  {\cal C}$ can be
obtained through the calculation of the resolvent \cite{Tho95}
\begin{equation}
G^{\cal C} _{ab} ( \lambda + i \epsilon) = \bigg( (\lambda + i\epsilon
) {\cal I}- {\cal D} ^  {\cal C} \bigg) ^{-1} _{ab} =
\sum _{e=1} ^N \frac{ w ^{\cal C}_{a,e} w ^{\cal C}_{b,e} }{ \lambda -
\lambda ^{\cal C}_e + i
\epsilon } \qquad ,
\label{fieldd}
\end{equation}
where ${\cal I}$ denotes the identity operator.

Introducing the density of eigenvalues 
\begin{equation}
\rho ^{\cal C} (\lambda )=\frac{1}{N} \sum _{e=1}^N
\delta ( \lambda - \lambda ^{\cal C}_e ) \qquad ,
\end{equation}
we may rewrite the discrete sum over eigenstates $e$ in (\ref{fieldd})
as an integral over eigenvalues with measure $\rho ^{\cal C}$.
The knowledge of the trace of the resolvent,
\begin{equation}
\frac 1N \sum _{a=1}^N G^{\cal C} _{aa}(\lambda + i \epsilon )  =
\frac 1N \sum _{e=1} ^N \frac{1 }{ \lambda - \lambda ^{\cal C}_e + i \epsilon }
= \int _{-\infty} ^\infty d\mu \; \frac{\rho ^{\cal C} (\mu ) }
{\lambda - \mu + i \epsilon } \quad ,
\label{poiu}
\end{equation}
gives then access to the density of states  through identity
\begin{equation}
\rho ^{\cal C} (\lambda )=  - \frac{1}{\pi} \lim 
_{\epsilon \to 0 ^+} \hbox{\rm Im} \left[ \frac 1N \sum _{a=1}^N 
 G^{\cal C} _{aa}(\lambda + i \epsilon )  \right] \qquad .
\label{spectre}
\end{equation}
Another quantity of interest is the autocorrelation function of
eigenvectors at distance $d$,
\begin{equation} \label{corr73}
A_e ^{\cal C} (d) = \sum _{a=1} ^{N-d} w^{\cal
C}_{a,e} w^{\cal C}_{a+d,e} \qquad . 
\end{equation}
We then define $A^{\cal C} (\lambda ,d) $ as the average value
of $A_e ^{\cal C} (d) $ over all eigenvectors $e$ lying in the 
range $\lambda \le \lambda ^{\cal C}_e \le \lambda + d\lambda$.
This autocorrelation function is simply related to the
off-diagonal resolvent (\ref{fieldd}) through
\begin{equation}
\frac{1}{N} \sum _{a=1}^{N-d} G^{\cal C} _{a,a+d}(\lambda + i \epsilon )
 = \int _{-\infty} ^\infty d\mu \; 
\frac{\rho ^{\cal C} (\mu )} {\lambda-\mu + i\epsilon} \; A^{\cal C} (\mu ,d)
\qquad ,
\label{rel1}
\end{equation}
that generalizes equation (\ref{poiu}) to non zero values of $d$. 
Taking the imaginary part of equation (\ref{rel1}), we obtain
\begin{equation}
\rho^{\cal C}(\lambda) \;A^{\cal C} (\lambda ,d) = 
- \frac{1}{\pi} \lim 
_{\epsilon \to 0 ^+} \hbox{\rm Im} \left[ \frac 1N \sum _{a=1}^{N-d} 
 G^{\cal C} _{a,a+d}(\lambda + i \epsilon )  \right] \qquad .
\label{auto}
\end{equation}
Therefore, the calculation of the large distance $d$ behaviour of
$G^{\cal C} _{a,a+d}  (\lambda + i \epsilon )$ 
will give access to the asymptotic scaling of the 
autocorrelation function 
\begin{equation} \label{asympto}
A^{\cal C} (\lambda ,d) \propto  \left( e^{- \sigma ^{\cal C} 
(\lambda )  + i \; q ^{\cal C} (\lambda ) }\right) ^d \qquad ,
\end{equation}
and thus to some effective relation of dispersion $\lambda ^{\cal
C}(q)$ and to the coherence length $1/\sigma ^{\cal C} (\lambda )$ 
of the instantaneous modes at finite temperature. 

\subsection{Average over the instantaneous molecular configuration}

To perform the average over the molecule configurations ${\cal C}$, 
we first rewrite the resolvent (\ref{fieldd}) 
as the propagator of a replicated Gaussian field theory 
\cite{Tho95}
\begin{eqnarray}
G^{\cal C} _{ab}(\lambda + i \epsilon )  &=& \frac{ -i 
\int \prod _{j=1}^N d x _{j} 
(x_a \, x _b )\; 
\exp \left(\frac{i}{2}( \lambda + i\epsilon )
\sum _{j=1}^N  x _{j} ^{\; 2} - \frac{i}{2} \sum _{j,k=1}^N 
{\cal D}^  {\cal C} _{j,k} \; {x}_{j}\, {x}_{k}  \right) }
{\int \prod _{j=1}^N d x _{j} \; 
\exp \left(\frac{i}{2}( \lambda + i\epsilon )
\sum _{j=1}^N  x _{j} ^{\; 2} - \frac{i}{2} \sum _{jk=1}^N 
{\cal D}^  {\cal C} _{jk} \; {x}_{j}\, {x}_{k}  \right) }
\nonumber \\
&=& \lim_{n\rightarrow 0}
- \frac in  \int \prod _{j=1}^N d\vec x _{j} 
(\vec x  _a .\vec x _b )\; 
\exp \left(\frac{i}{2}( \lambda + i\epsilon )
\sum _{j=1}^N \vec x _{j} ^{\; 2} - \frac{i}{2} \sum _{j,k=1}^N 
{\cal D}^  {\cal C} _{jk} \; \vec {x}_{j}.\vec {x}_{k}  \right)
 \quad
\label{field}
\end{eqnarray}
Replicated fields $\vec x _j = (x_j ^1 , \ldots , x_j ^n )$ 
are $n$-dimensional vector fields attached to each site $j$. 
The positivity of $\epsilon$ ensures that the Gaussian integrals in
(\ref{field}) are well defined. 

At equilibrium, the molecular configuration ${\cal C}$ is drawn 
according to the Gibbs measure (\ref{gibbs}).
We shall denote the average of any quantity over distribution
(\ref{gibbs}) in the same way as its configuration dependent counterpart 
but without ${\cal C}$ subscript. For instance, the average resolvent
reads
\begin{equation}
G _{ab}(\lambda + i \epsilon ) = \int _{r_{min}} ^\infty r_1 dr_1 \int
_{-\infty} ^\infty d\varphi _1 \ldots \int _{r_{min}} ^\infty r_N dr_N
\int _{-\infty} ^\infty d\varphi _N \; {\cal P} ( \{ r_j ,\varphi _j
\}) \ G ^{\{r_j, \varphi _j \}} _{ab}(\lambda + i \epsilon ) \ ,
\label{fieldmoyen}
\end{equation}
where ${\cal P}$ is the Gibbs measure (\ref{gibbs}).
It is important to keep in mind that ${\cal D} ^{\cal C} _{jk}$
vanishes for $|j-k|\ge 2$, see Section III.2. Therefore, only
replicated variables $\vec x_j , \vec x_k$   corrsponding to
adjacent base pairs along the molecule ($j=k \pm 1$) interact
together in the expression (\ref{field}) of the resolvent. 

\subsection{Transfer matrix formalism}

The one dimensional structure of the interactions in
(\ref{fieldmoyen}) can be exploited through the introduction of a
transfer matrix ${\cal T}$ relating the molecular variables $r_j ,
\varphi _j$ as well as the replicated variables $\vec x_j$ to their
counterparts at site $j-1$. The entries of this matrix can be read
from (\ref{hamil},\ref{field},\ref{fieldmoyen}),
\begin{eqnarray}
\langle r,\varphi , \vec x |{\cal T} |r',\varphi' , \vec x' 
\rangle = T(r,r',\theta )\; \exp \bigg(  &&  
\frac i4 (\lambda + i \epsilon ) ( \vec x^2 + \vec x'^2 )
- \frac i4 d_0 (r,r',\theta )\vec x^2  \nonumber \\
&& \left.  - \frac i4 d_0 (r',r,-\theta
 )  \vec x'^2  + \frac i2 \;
d_1 (r,r',\theta ) \vec x . \vec x' \right) \quad . \label{tr2}
\end{eqnarray}
In the above expression, $d_0$ and $d_1$ are respectively the diagonal
and off-diagonal elements of ${\cal D}^{\cal C}$, see (\ref{Hess})
while $T$ has been defined in (\ref{deft}). Notice that ${\cal T}$ is
a symmetric (but not Hermitian) matrix.

By definition, the average resolvent $G_{ab}$ is the
correlation function of replicated variables $\vec x_a$ and $\vec x_b$
(\ref{field}). Within the transfer matrix formalism, this mean dot 
product can be computed from the knowledge of eigenvalues 
$\Lambda _\ell$ and all (normalized) eigenvectors $\Psi_\ell 
(r,\varphi , \vec x )$ of ${\cal T}$. Calling $d$ the axial distance 
$|a-b|$ between  sites $a$ and $b$ along the molecule,  
the average resolvent reads in the thermodynamical limit
\begin{equation} 
\label{corre}
G_{ab}(\lambda + i \epsilon ) 
 =  \lim_{n \rightarrow 0} - {i}\sum _{\ell = 0}^\infty 
\left( \frac{\Lambda _\ell}{\Lambda _0} \right)^{d}
\left[ \int d\vec{x}\; dr\; d\varphi\; x^1\; \Psi_0 
(r,\varphi , \vec x ) \; \Psi_\ell 
(r,\varphi , \vec x )   \right]^{2}
\end{equation}
The maximal eigenvalue (in modulus) of ${\cal
T}$ coincides with $\ell =0$ and increasingly excited states
correspond to $\ell \ge 1$. 

The average diagonal resolvent $G_{aa}$ and thereby the mean 
density of states may be obtained from (\ref{corre}) when $d=0$, 
\begin{equation} \label{spec2}
\rho (\lambda ) = \frac 1\pi \lim _{\epsilon \to 0^+ , n \to 0}  
\hbox{Re} \left[\int d\vec{x}\; dr\;d\varphi \;
\Psi_0(r,\varphi , \vec x )^{2} \;  (x^1)^{2}\right] \quad .
\end{equation}

Expression (\ref{corre}) may also be used to compute the
autocorrelation function of eigenvectors at large distances $d$.
for which the resolvent (\ref{fieldmoyen}) scale asymptotically as, 
see (\ref{asympto}), 
\begin{equation}
G_{a,a+d} (\lambda + i\epsilon ) \propto \exp (- \sigma (\lambda )\; d +
i \; q (\lambda) \; d ) \qquad ,
\label{rel2}
\end{equation}
where both $\sigma (\ge 0)$ and $q$ depend on the
energy level $\lambda$. When $d$ is large, the sum in (\ref{corre}) 
is dominated by the $\ell=1$ contribution and equation (\ref{rel2})
may be reformulated as
\begin{eqnarray}
\lim _{d\to \infty } \; \frac
1d \; \ln \; G_{a,a+d} (\lambda + i\epsilon ) 
&=&- \sigma (\lambda ) + i \; q (\lambda ) \label{rel24} \\
&=& \ln \left( \frac{ \Lambda _1}{\Lambda _0} \right) 
\qquad , \label{rel25}
\end{eqnarray}
which allows to derive $\sigma$ and $q$ from the knowledge of
$\Lambda_0$ and $\Lambda _1$.  We shall explicitely compute the
inverse length $\sigma$ and wave number $q$ in Section V and compare
them to the zero temperature results of Section II.D.

Notice that $\sigma$ and $q$ 
defined in (\ref{rel2}) do not exactly coincide with the 
values of $\sigma ^{\cal C}$ and $q^{\cal C}$ appearing in (\ref{asympto})
averaged over ${\cal C}$ with distribution (\ref{gibbs}).
To obtain the latter, one should average the logarithm of the
resolvent $G^{\cal C}$ (quenched average) and not simply 
compute the logarithm of the mean resolvent as in (\ref{rel24})
(annealed average) \cite{luck}. Due to concavity
of the logarithm function, $\sigma$ defined in (\ref{rel24}) is not
only a approximation but also a lower bound to the average value 
of $\sigma ^{\cal C}$. Comparison with numerical simulations made 
for the so-called ``smallworld''  lattice have shown that $\sigma$
provides a very good estimate of the mean $\sigma ^{\cal
C}$\cite{Mon99}. To sum up, we can compute from the average
resolvent $G$ an upper bound $1/\sigma$ to the coherence length 
of the eigenvectors at level $\lambda$ as well as an estimate $q$ of
their typical wave number.

\subsection{Rayleigh-Ritz formula and variational approach}

The diagonalization of the transfer matrix and the analytical
continuation in $n\to 0$ could in principle be performed exactly due
to the rotational invariance of ${\cal T}$ in the $n$-dimensional
space of replicated variables. To avoid this tedious calculation, we
resort to a Gaussian variational approach whose accuracy and
reliability has been recently validated in the case of
diffusion on random lattices \cite{Bir99,Mon99}. 

We start from the Rayleigh-Ritz formula for the largest eigenvalue
$\Lambda _0$ of a (real-valued) matrix ${\cal T}$,
\begin{equation}
\Lambda _0 = \max _{\Psi \ne 0 } {\cal R} ( \Psi ) \qquad,\label{rr}
\end{equation}
where
\begin{equation}
{\cal R} ( \Psi ) = \frac{ \langle \Psi | {\cal T} | \Psi  \rangle}
{\langle \Psi | \Psi \rangle } \qquad .
\label{rr2}
\end{equation}
A lower bound $\Lambda _0 ^g$ to $\Lambda _0$ can be obtained through 
maximization of the r.h.s. of (\ref{rr}) within a suitable trial
family of wave functions $\Psi _0 ^g (Q)$ parametrized by some tunable 
variables $Q$. According to (\ref{rr}, \ref{rr2}), we obtain
\begin{equation} \label{rr3}
\Lambda_0 ^g = {\cal R} \big(\Psi _0 ^g (Q_{opt}) \big) \qquad ,
\end{equation}
where the optimal parameter $Q_{opt}$ is the solution of 
\begin{equation} \label{rr4}
\left. \frac{d {\cal R} \big(\Psi _0 ^g (Q) \big)}{dQ} \right| 
_{Q=Q_{opt}} =0 \qquad .
\end{equation}
Notice that within formulation (\ref{rr3},\ref{rr4}),  ${\cal T}$
needs no longer to be real-valued and can assume complex values.

Hereafter, we  apply this procedure to the transfer matrix (\ref{tr2}) and
resort to the following Ansatz,
\begin{equation}
\Psi _0 ^g (r,\varphi , \vec x ) = \psi _0 (r) \;  \exp \left(
\frac i4 \; Q(r) \; \vec x ^2 \right)
\qquad , \label{trwf}
\end{equation}
where $\psi_0 (r) $ is the maximal wave function of the $T_0$ transfer
matrix (\ref{autremu}) of Section II.B. This choice ensures that the correct 
eigenvalue $\lambda _0$ is recovered when the number of replicas $n$
strictly equals zero: $\Lambda _0 (n) = \lambda _0 + O(n)$. The $\vec
x$ dependence of the trial wave function (\ref{trwf})results from
the similarity of the operator ${\cal T}$  
(\ref{tr2}) with the transfer matrix of a 1D chain of interacting
spherical spins, for which the ground state wave function is Gaussian
 \cite{parisi}. Furthermore, this Ansatz allows for 
an exact calculation of the $n$-dimensional integral over $\vec x$ and 
an easy continuation of the result to real values of $n$.
Note here that the wave function (\ref{trwf}) does not depend on
$\varphi$. This is no assumption for $n=0$, see Section II.B and supposed
to give quantitatively good results for $n>0$.

The calculation of the Rayleigh-Ritz functional ${\cal R}$ (\ref{rr2})
with the Gaussian Ansatz (\ref{trwf}) is exposed in Appendix A. The
resulting functional optimization equation over $Q$ reads
\begin{equation}
\frac {\lambda _0 \;\psi _0 (r)}{2\; Q(r)} = \int _{r_{min}} ^\infty
 dr' \int _{0} ^\pi d\theta \;  T(r,r',\theta )\; 
\psi _0 (r') \; \frac{  Q(r') + \lambda + i\epsilon - 
d_0 (r',r,-\theta )}{ {\cal W} (r,r',\theta )} \quad ,
\label{opt1}
\end{equation}
where 
\begin{equation}
{\cal W} (r,r',\theta ) =
(Q(r) + \lambda + i\epsilon -  d_0 (r,r',\theta ) ) 
(Q(r') + \lambda + i\epsilon -  d_0 (r',r,-\theta ) ) 
- d_1 (r,r',\theta ) ^2\quad .
\label{opt16}
\end{equation}
The solution of the above equation gives access to the variational
estimate of the density of modes from (\ref{spec2}),
\begin{equation}
\rho ^g (\lambda ) = -\frac 1\pi \int _{r_{min}} ^\infty
dr\; \psi _0 (r) ^2 \; \hbox{\rm Im} \left( \frac 1{Q(r)} \right)
\qquad . \label{spec87}
\end{equation}

We can pursue the procedure to compute the  excited eigenvalues
$\Lambda _\ell$  of ${\cal T}$. Drawing our inspiration from the
1D spherical spins chain transfer matrix \cite{parisi}, we look for a
variational wave function of the type
\begin{equation} \label{trwf1}
\Psi _\ell ^g (r,\varphi , \vec x ) =  \Psi _0 ^g (r,\varphi , \vec x )
\; P ^g _\ell ( x^1 , x^2 , \ldots , x^n ) \qquad ,
\end{equation}
where $P^g _\ell $ is a polynomial of $n$
variables of total degree $\ell$.  To obtain the asymptotic
behaviour of the resolvent (\ref{rel2}), we only need the first
excited state $\ell =1$, whose corresponding variational polynomial is
clearly $P^g _1 (\vec x ) = x^1$, giving thereby from (\ref{rr2})
\begin{eqnarray}
\tau ^g = \frac{\Lambda _1 ^g}{\Lambda _0 ^g} = -2\ && \left\{
\int _{r_{min}} ^\infty dr dr' \int _{0} ^\pi d\theta \; 
T(r,r',\theta) \;\psi _0 (r)\; \psi _0 (r')\; \frac{d_1 (r,r',\theta ) }{
{\cal W} (r,r',\theta )} \right\} \nonumber \\
&& \times  \left\{ 
{\lambda _0 \int _{r_{min}} ^\infty dr\; \psi _0 (r) ^2 Q(r)^{-1} } \right\}
^{-1} \label{lam1}
\end{eqnarray}
in the $n\to 0$ limit as shown in Appendix A.

%
%

\section{Results and Comparison with Experiments}

\subsection{Numerical resolution of the self-consistent equations}

We first compute the ground state wave function $\psi _0 (r)$ and the
corresponding eigenvalue $\lambda _0$, see Section
II.B, by means of Kellog's method \cite{cocco}. The integration range
$[r_{min};r_{max}]$ over $r$ is discretized 
 into a set of $n_r$ points $r_\alpha$, $\alpha=1,\ldots , n_r$ with $r_1 =
r_{min}=9.8 {\AA}$ and $r_{n_r} = r_{max}=10.7 {\AA}$. 
  
Self-consistent equations (\ref{opt1}) for $Q(r)$ can then be solved
iteratively. As can be checked at zero temperature, convergence is improved 
by iterating the equations for $1/Q(r)$ rather than for $Q(r)$ itself. 
We stop the iteration process as soon as the differences between the 
$1/Q(r_\alpha)$'s and their images through the iteration become smaller 
than $10^{-7}$ for all $1\le \alpha \le n_r$. 

Numerical difficulties come from the limits
$\epsilon \to 0$ and $n_r \to \infty$. This can be best seen for base
pairs vibrations in the simplest case $E=0$. The exact solution to
equations (\ref{opt1}) is $Q(r)=\lambda + i\epsilon - 
\frac 12 V''_m (r)$. In other
words, eigenvalues $\lambda $ and radii $r$ are in simple
correspondance~: to any permitted eigenvalue $\lambda$ is associated one 
(or a few) radius $r(\lambda )$ such that $\lambda =  \frac 12
V''_m (r(\lambda ))$ and the density of states reads
\begin{equation} \label{nume1}
\rho (\lambda ) = \lim _{\epsilon \to 0}
\frac 1\pi \int _{r_{min}} ^{r_{max}} dr  \frac{ \psi _0 (r) ^2 \;
\epsilon}{ (\lambda - \frac 12 V''_m (r))^2 + \epsilon ^2}
= \frac{2}{| V'''(r(\lambda ))|}\; \psi _0 (r (\lambda )) ^2 \quad .
\end{equation}
In practice however, the integral in (\ref{nume1}) is discretized as
follows
\begin{equation} \label{nume2}
\rho (\lambda ) = \lim _{\epsilon \to 0} \frac 1\pi
\sum _{\alpha =1} ^{n_r} \frac { \psi _0 (r_\alpha ) ^2 \;
\epsilon}{ (\lambda - \frac 12 V''_m (r_\alpha))^2 + \epsilon ^2}
\qquad .
\end{equation}
Consider the eigenvalue $\lambda _\gamma$ corresponding to radius
$r_\gamma$ for some arbitrary integer $\gamma$. Dominant
$O(1/\epsilon) $ contributions to the density $\rho (\lambda _\gamma
)$ in (\ref{nume2}) come from $\alpha$ in the range $\gamma -\Delta
\le \alpha \le \gamma + \Delta $ with $\Delta \simeq2 \; n_r \; \epsilon /
V'''(r_\gamma )$. Problems arise when $\Delta$ is close to unity. When
$\lambda $ scans the interval $[\lambda _\gamma ; \lambda _{\gamma
+1}]$, the index of the only (for $\Delta =1$) contributing term to
$\rho (\lambda )$ jumps from $\alpha =\gamma$ to $\alpha =\gamma +1$
at some intermediate $\lambda$ which will be a local minimum of the
density $\rho $. Such local fluctuations are pure artifacts of the
discretization procedure and must be removed by keeping $\Delta \gg 
1$, that is $1/n_r \ll \epsilon \ll 1$.  Typical suitable values of the
parameters are $\epsilon = 2.10^{-2}$, $n_r = 4.10^3$, giving a spectrum
$\rho (\lambda )$ almost normalized to unity (with a small error
$\simeq \epsilon $). 

Note that the same reasoning holds for the
numerical calculation of the torsional spectrum. The discretization of
the integral over $0<\theta <\pi$ must be replaced by a sum involving a large
number $n_\theta =900$ of terms to reach a good accuracy of the
results. 

\subsection{Radial modes}

Radial mode spectra obtained for different choices of $D$ and $E$ are
displayed figure~5. They exhibit smooth shapes and
Van Hove divergences have been smeared out by thermal disorder,
compare to figure~4. As at zero temperature, the overall
width of the spectrum is an increasing function of $E$.

For $E=4 eV/{\AA}^2$, the general form of $\rho (\nu )$ is reminiscent of the
density of states at zero temperature, with a shoulder in $\nu_-
\simeq 75$~cm$^{-1}$ and a maximum in $\nu_+ \simeq 135$~cm$^{-1}$, in
correspondence  with the edges of the zero temperature
spectrum, $\nu_- ^0 = 73.8$~cm$^{-1}$ and  $\nu_+ ^0 
= 138.7$~cm$^{-1}$. A careful analysis even show a quantitative
agreement between the density of states at $T=300$~K and $T=0$~K for
frequencies lying in the range 110~cm$^{-1} < \nu < 130$~cm$^{-1}$. 
A similar behaviour, that is the robustness of the central part of the
spectrum to (weak) disorder was also observed in~\cite{Mon99}.

At a weaker stiffness $E=0.74 eV/{\AA}^2$, a single bump is
observed. The range of allowed frequencies at zero temperatures $\nu_-
^0 = 71.5$~cm$^{-1} < \nu < 87.5 $~cm$^{-1}$ is indeed too narrow and
both peaks merge under the action of thermal disorder. Note that a
very small fraction of modes seem to be unstable and give rise to imaginary
frequencies, see Section~III.B. We however discard them since their 
integrated sum is smaller than the accuracy $\epsilon $ of the calculation.

Figure~6 shows the dispersion relations at ambient
temperature for radial modes. The frequency $\nu _r (q)$ is an
increasing function of the wave number over the interval $0\le q\le
\pi$. Since the range of allowed frequencies is much larger at $T=300$~K
than at $T=0$~K, there is no general coincidence with the zero
temperature dispersion relations as can be seen for $E=0.74 eV/{\AA}^2$.
For larger stiffness constants, the dispersion relations for
both temperatures however coincide for medium wave numbers, 
{\em i.e.} when $1\le q \le 2$ roughly. In agreement with the
above analysis of the density of states, the effective thermal
disorder gets weaker and weaker as the stiffness constant $E$ grows. 

When $E$ increases, the molecule becomes more and more rigid since
radii $r_j$ less and less fluctuate from base pair to base pair. The
wave function $\psi_0 (r)$ gets more and more concentrated around the
minimum of the Morse potential $r=R$, see figure~2 and the
region of integration over $r$ that mostly contributes to the density
of states in (\ref{spec87}) becomes narrower and narrower. On the
opposite, for small $E$, $\psi _0 (r)$ mainly reflects the structure
of the Morse well whose flanks are not accessible at zero temperature. 
The tails of $\psi  _0$ are large and give rise to some tails for the
density of states. The cross-over between both regimes takes place at
$E \sim D\; a^2$, that is of the order of a few $eV/{\AA}^2$. 

The inverse coherence length $\sigma _r$ is plotted
figure~7 as a function of frequency. In the central
ragion of the spectra, the corresponding autocorrelation lengths are
$\xi \simeq 0.7$ for $E=0.74$ and $\xi \simeq 4$ for $E=4$ with a more
sensitive dependence on $\nu$ in the latter case. As expected from the
above discussion, $\xi$ increases with $E$. 
The values of the coherence length $\xi$ are in good agreement with 
the equilibrium correlation distance $\zeta$ defined through
\begin{equation}
\big\langle (r_j - \langle r\rangle ) ( r_{j+d} - \langle r\rangle
) \big\rangle \propto e^{-  d / \zeta} 
\qquad \qquad (d\to \infty ) \qquad .
\end{equation}
A thermodynamical calculation of $\zeta$ can be carried out from 
the knowledge of the excited states of the transfer matrix $T_0$
confined to the Morse potential \cite{futur}. Results are
$\zeta \simeq 0.51$ for $E=0.74$ and $\zeta \simeq 0.93$ for
$E=4$\cite{futur}. Note that $\zeta$ is the inverse damping with of
the static structure factor whereas $\xi (\nu)$ is an energy (frequency)
dependent coherence length.

\subsection{Angular modes}

We now turn to the angular spectrum. The density of modes is shown on
figure~8. We first concentrate on positive, that is
real frequencies. The band edge $\nu_+ ^0  = 13.6 $~cm$^{-1}$  
of the zero temperature spectrum visible on figure~4 
disappears at finite temperature. A sharp maximum now takes place at
$\nu_M \simeq 7$~cm$^{-1}$. As expected from zero temperature
calculations, the width of the peak is smaller than for radial modes
and can be estimated to $\Delta \nu \simeq 15 $~cm$^{-1}$.

The densities of states at $E=0.74$ and $E=4$ coincide within $0.1
\%$. We have numerically checked that the angular mode spectrum
depends extremely weakly on the stiffness constant $E$ over the whole
range $0\le E\le \infty$. In other words, unlike radial modes, angular
modes are not sensitive to the width of the ground state wave function
$\psi _0$, {\em i.e.} to fluctuations of the base pair radius $r$, see
figure~2. This observation is supported by inspection of the
variational parameter $Q(r)$ entering wave function~(\ref{trwf}). At
fixed frequency, both imaginary and real parts of $Q(r)$ are indeed
almost constant on the whole range of radius $9.8 {\AA}< r < 10.7 {\AA}$.

The robustness of the spectrum displayed fig.~8 can be
understood by looking at the variations of $d_0 (r,r',\theta )$ around
the thermal average positions $r=r'=\langle r_n \rangle$ and $\theta = \langle
\theta_n \rangle$. The second derivative of $d_0$ with respect to $r$ 
at fixed twist angle equals $\partial ^2 _r d_0 \simeq 0.045 eV/{\AA}^4$
while the range of fluctuations of $r$ is given by the largest squared
width of $\psi _0$ (corresponding to $E=0$) and reads $\Delta \langle r_n 
^2 \rangle \simeq 0.04 {\AA}^2$. Therefore the fluctuations of the
radius $r$ modifies $d_0$ by  $\frac 12 \partial ^2 _r d_0
. \Delta \langle r_n ^2 \rangle \simeq 0.001 eV/{\AA}^2$, that is
by less than $2\%$ of $d_0$ typically. Repeating the same calculation for
twist-induced fluctuations, we obtain $ \partial ^2 _\theta d_0 
\simeq 20 eV/{\AA}^2$ and $\Delta \langle \theta _n 
^2 \rangle \simeq k_B T/ (d_0  R^2) \simeq 0.005 \ rad^{\, 2}$. The resulting
variations of $d_0$ due to changes of twist are of the order of
$\frac 12 \partial ^2 _\theta d_0 . \Delta \langle \theta _n ^2 \rangle \simeq
0.05 eV/{\AA}^2$ that is comparable to $d_0$. 

Consequently, an excellent approximation of the angular mode spectrum 
may be obtained by the following simple argument. Let us call 
\begin{equation}
d_0 (\theta ) = \frac 1{R^2} \;\frac{d^2 V_b}{d\theta ^2}
(R,R,\theta ) \label{dt0}
\end{equation} 
the diagonal element of ${\cal D} ^{\cal C}$ (\ref{Hess}) where we have
for simplicity identified $\langle r_n \rangle$ with $R$ since
$\langle r_n \rangle - R \simeq 0.01 {\AA}\ll R$ at $T=300$~K). 
The twist angle $\theta$ is approximately distributed with the Gibbs
measure, see (\ref{mu}),
\begin{equation}
p ( \theta ) = \frac 1{ Y_0(R,R)} \;
  \exp \bigg\{ -\beta  
V_b\,(R,R,\theta)  \bigg\} \quad . \label{dt1}
\end{equation} 
Then we may substitute the variational equation (\ref{opt1}) on $Q(r)$ with 
\begin{equation}
\frac {1}{2\; Q} = \int _{0} ^\pi d\theta \, p(\theta )\, 
\frac{ Q + \lambda + i\epsilon - 
d_0 (\theta )}{(Q + \lambda + i\epsilon -  d _0(\theta ) ) ^2
- d_0 (\theta ) ^2}\ ,
\label{opt10}
\end{equation}
which involves a single parameter $Q$. Solving equation (\ref{opt10}),
the density of angular modes equals $-$Im($1/Q$)$/\pi$ and is in
excellent agreement with figure~8. From a numerical
point of view, this approximation is much less time consuming than the
full resolution of (\ref{opt1}). In fact, equation (\ref{opt1}) required 
the integral over $\theta$ to be computed for each value of $r$ (and at
each iteration step) and the solving time was therefore $n_r$ times larger
than for (\ref{opt10}).

Close to zero frequency, the density of states vanishes as $\rho
_\varphi (\nu ) \propto |\nu |$ since the density of eigenvalues $\rho
(\lambda )$ is finite in $\lambda =0$, see (\ref{ris}).
We now turn to negative, that is imaginary frequencies. Computing the 
integrated density of unstable modes, we see that the latters
represent roughly $20\%$ of angular modes. They extend down to
frequencies equal to $\nu _- =-20$~cm$^{-1}$ with a maximum in $\nu \simeq
-3.5$~cm$^{-1}$. We shall come back to the physical implications of 
these modes in next Section.

The relation of dispersion $\nu _\varphi (q )$ for the angular modes
is displayed figure~9 over the whole range of real and
imaginary frequencies. We also show on figure~10 the
inverse autocorrelation length $\sigma _\varphi (\nu )$. As for radial
modes, the dispersion relations at $T=300$~K gets close to its zero
temperature counterpart at intermediate wave numbers $0.5 \le q \le
1.5$ corresponding to frequencies $0$~cm$^{-1} \le \nu \le
10$~cm$^{-1}$. This coincidence is accompagnied by small values of
$\sigma$ on this interval of frequencies, giving rise to a
coherence length of the order of $\xi \simeq 2$. 
Conversely, large frequencies correspond to highly disordered modes:
$q\ge 2$ and monotonously increasing $\sigma $, with $\xi \simeq 0.4$
at $\nu =30$~cm$^{-1}$. Notice that the statical correlation length is
the same as for radial modes, see Section~V.B.

Unstable modes have also short autocorrelation lengths, e.g. $\xi 
\simeq 0.4$ at $\nu =-20$~cm$^{-1}$. However, their wave numbers are much
smaller and can be considered as contant (and zero) for $\nu \le
-10$. The autocorrelation function of unstable eigenmodes
(\ref{corr73}) therefore do not change sign over a typical distance $d_s
\sim \pi/q \gg 1$.  Unstable modes can be seen as unstable
acoustic phonons, involving coherent rotations of the molecule
extending over regions of size $d_s$. 

To end with, notice that the dispersion relation and the inverse coherence
length both exhibit an inflection point in $\nu =0$ as shown
fig.~9~and~10. This is an artefact of the
representation of unstable modes as negative frequencies. 
Consider for instance the damping width $\sigma$ close to zero
frequency. In the natural $\lambda$ eigenvalue parametrization, we
expect a non singular behaviour in the vicinity of $\lambda =0$:
$\sigma (\lambda ) = \sigma _0 + \sigma _1 \lambda + O(\lambda
^2)$. For positive eigenvalues $\lambda$, the frequency $\nu$ is
defined as $\nu \propto \sqrt \lambda $ (\ref{eq}) while using the
negative-imaginary convention of Section~III.B, $\nu \propto - \sqrt {-
\lambda }$ for negative eigenvalues. The expansion of the inverse
coherence length as a function of (small) frequencies thus reads:
$\sigma (\nu ) = \sigma _0 + \tilde \sigma _1 \nu . |\nu | + O(|\nu |^3)$ and
is singular with an inflection point in $\nu=0$. 

\subsection{Discussion}

Our calculation shows that the coherence length of normal modes is
finite and remains of the order of unity. Eigenvectors at finite
temperature are thus far from being plane waves as in the zero
temperature case. The power spectrum, that is the Fourier transform of
the autocorrelation function at frequency $\nu$ (\ref{rel2}) acquires
a Lorentzian form centered around a certain wave number $q(\nu )$ with
a width $\sigma (\nu )$. This behaviour is experimentally observed in
neutron scattering experiments~\cite{Gri87,Gri88}. Furthermore,
the calculation justifies {\em a posteriori} the absence of selection
rule on momentum in Raman experiments.  Indeed, the coherence length
of the disordered phonons is of the order of $\xi H \le 20 {\AA}$ and is
negligible with respect to optical wavelengths $\simeq 4800 {\AA}$, see
Section~III.C.

Both radial and angular spectra are superposed in figure~11.
The total spectrum $\rho (\nu)$ equals $\rho _r(\nu) + \rho _\varphi
(\nu )$ is normalized to unity (half of modes originate from angular
degree of freedom, and the remaining half from radial ones).  
The torsion peak appears much more narrow and higher than its radial
counterpart. The width of the latter amounts to $\Delta \nu \simeq 
60$~cm$^{-1}$ (respectively $\Delta \nu \simeq 
100$~cm$^{-1}$) for $E=0.74 eV/{\AA}^2$ (resp. for $E=4 eV/{\AA}^2$) 
whereas (stable) torsional modes spread over a range of
$\Delta \nu \simeq 15$~cm$^{-1}$. While for $E=4 eV/{\AA}^2$, torsional
and radial spectra do not intersect each other, there is an overlap
region at smaller stiffness constant around $\nu \simeq
30$~cm$^{-1}$. Nevertheless, both fluctuations take place on basically
two distinct scale times. Angular vibrations are associated to a
typical frequency of $\simeq 10$~cm$^{-1}$, or equivalently to a
typical time $\tau _v \simeq 3\; 10^{-12}$~s. Radial modes are present
at frequencies $\simeq 100$~cm$^{-1}$ that is involve dynamical
processes on a scale $\tau _r \simeq 0.3\; 10^{-12}$~s. At low
frequencies $|\nu|< 25$~cm$^{-1}$, the presence of unstable modes
threatens the validity of the linearization procedure used in
Section~III.B. A more refined analysis taking into account non
linear terms in the dynamical equations is needed to understand how
unstable modes (with low wave numbers $q$) are coupled to stable mode
(with larger $q$) and modify the frequencies of the
latters. the influence of viscous friction upon modes in this frequency
regime would also deserve to be further studied as mentioned
Section~III.A. Conversely, the absence of unstable modes at frequencies
larger than $\nu > 25$~cm$^{-1}$ indicate that the INM predictions are
reliable for time scale smaller than $10^{-12}$~s. This is enough to
identify the range of allowed frequencies for torsional modes: $\nu <
30$~cm$^{-1}$. The INM prediction for the location of the torsional 
peak $\nu _M \simeq 7$~cm$^{-1}$ cannot be trust blindly but it
reasonably lies in the middle of the zero temperature band, see
Section~III.D.

We now turn to the comparison with spectroscopy measurements. To
establish the link between the density of states and the Raman
intensity, the knowledge of the light-to-vibrations coupling $C(\nu)$
is necessary as seen Section~III.C. In the absence of precise
information on the latter, we have rescaled the density of modes $\rho
(\nu)$ according to formula (\ref{rescaraman}) for the three different
choices $C(\nu )=1$, $C(\nu )=\nu$ and $C(\nu )=\nu ^2$ (note that the
second hypothesis for $C$ is the most plausible one).  The resulting
theoretical Raman intensities are shown figure~12.a for
$E=0.74 eV/{\AA}^2$ and figure~12.b for $E=4 eV/{\AA}^2$. From the
one hand, the overall shape of the spectra change with $C$ with
respect to $\rho$. In particular for $C(\nu) =1$ and $C(\nu )=\nu$,
the torsional peak diverges at small frequencies and the radial modes
bump acquires a shoulder from, located in the right flank of the
torsional modes. On the other hand, the band of allowed frequencies
remains unaltered by the choice of $C$ and extends over $50 < \nu <
110$~cm$^{-1}$ for $E=0.74$ and $50 < \nu < 150$~cm$^{-1}$ for
$E=4$. In the latter case, the radial region of the spectrum exhibit
two local maxima at the same height for $C(\nu )=\nu$. The rescaled
spectrum of fig~12.a for $C(\nu ) =\nu$ closely agrees with
intensity curves obtained from Raman experiments at $T=50^o$C and shown
figure~2 of Ref.\cite{Ura81} on the range of frequencies $50 < \nu <
110$~cm$^{-1}$. The latter measurement was performed on calf thymus in
solution (10 mM PHB, ph=7). This allows us to think that $E=0.74$ is a
better choice for the stiffness constant than $E=4$~\cite{cocco}.

Our results do not show drastic variations with temperature for $T$
ranging from 300~K to the melting temperature $T_m = 350$~K. At $T_m$
the double helix structure disappears and so do all angular and 
radial vibrations. Since denaturation is described as a first order
phase transition by the present model, there is no gradual 
destabilization of the modes and no relevant change in $\rho$ can be
seen as mentioned above. The experimental Raman measurements
shown on figure~2 of Ref.\cite{Ura81} nevertheless show a smooth
change in the shape of the intensity curves over the temperature
interval $50^o$C$<T<80^o$C. This apparent paradox can be easily
explained by the fact that our model describes a homogeneous sequence
of bases. For the latter, the fraction of open base pairs versus
temperature exhibits an abrupt jump from zero to one at $T_m$~\cite{Saen84}.
In the case of a heterogeneous sequence as in the experiments of
Ref.\cite{Ura81}, the denaturation temperature of AT rich regions can
be inferred to be $T_m = 50^o$C~\cite{Saen84} and the fraction of open
base pairs smoothly increases from zero to one over the range of
temperatures $50^o$C$<T<77^o$C.  

We now turn to angular modes. Raman measurements on fibers at 100\% of
relative humidity (r.h.) have given evidence for a narrow band at
$16$cm$^{-1}$, see fig.~1 of Ref.\cite{Ura85}. This peak shifts down to
lower frequencies ($10$cm$^{-1}$) when increasing the hydration degree
of DNA in gels and disappears in the central component for DNA in
solution\cite{Ura81}. Other measurements on B-DNA fibers al lower
$\simeq 80$\% r.h. by Lindsay et al. have reported a similar band at a  higher
frequency $\nu \simeq 25$cm$^{-1}$ that shifts to lower values as the r.h.
increases in good agreement with Urabe et al.'s data~\cite{Lin88}.
The observed softening of the frequence may come
from the weakening of external e.g. interhelical interactions as well
as from the influence of the (rigid) primary and (viscous) secondary 
water shells, see Section~III.A. Note that this modes seems to appear when the
scattering vector $\Delta \vec Q$ gets parallel to the helical axis.
Other molecular system as crystals of ATP or guanosine monosphosphate 
(GMP) that present a columnar stacking of bases (even without hydrogen
bond interactions) exhibit similar low frequency spectra. The
intensity of the low frequency ($<20$cm$^{-1}$) mode was found to 
depend mostly on the stacking degree of bases in a column~\cite{Ura91}. 
Our calculation predicts, despite the presence of unstable INM and the
absence of friction terms in the dynamical equations (\ref{eqm}) that
angular modes are predominant for $\nu < 25$cm$^{-1}$. At such low
frequencies, radial modes can be neglected and twist fluctuations
are responsible for collective vibrations of interplane distances
$h_n$.  In agreement with experiments, our calculation thus
predicts that low frequency modes ($\nu < 25$cm$^{-1}$)
are related to vibrations of the base pairs column.

A quantitative comparison of our dispersion relations for angular
modes with experimental results is difficult to the lack of available
data. Neutron scattering measurements have shown that
pseudo-dispersion relations with a finite damping width can be
obtained for low frequencies but these experiments have been performed
on crystalline DNA fibers. Our model (and the values of parameters
exposed Section~II.C) are valid for DNA in solution where the
interaction with surrounding water is radically different and
interhelical effects are absent. More precisely, we find that at fixed
momentum $q$, the experimental frequency $\nu$ for fibers is 
larger than the theoretical predictions for diluted DNA. The
additional mass due to primary water shells for DNA in solution may
account for the reduction of frequency to a large extent~\cite{Lin88}.
Conversely, the extrapolated value of $\sigma$ is constant and equal
to 0.48 in good agreement with the theoretical minimum $\sigma =0.5$
shown fig.~10.

%
%

\section{Summary and Conclusion}

In this article, we have shown how to apply the INM framework to
a simple model of DNA molecule and reproduce the main features of
collective vibration modes at finite temperature.
Scale time separation between atomic vibrations and collective modes
allows to obtain good results at low frequencies without  resorting to
a detailed description of DNA at the atomic level. This simplified
modelization of DNA permits in turn a deeper analytical understanding 
of the structure of the modes than e.g. within MSPA. 

The model we had introduced to reproduce thermally and 
mechanically-induced DNA denaturation transitions 
has proven to be also capable of describing accurately the pico-second
dynamics seen through spectroscopy measurements. This robustness
has been obtained without any modification of the model or any new fit of
the constant force or geometrical parameters. Remarkably, the
comparison with Raman experiments has permitted us to decide the value
of the only parameter known with some uncertainty from the
denaturation experiments, that is the stacking stiffness $E$. 
The success of the present model to account for
completely different experiments comes from its mesoscopic nature,
that lies at an intermediate level between microscopic modelizations,
e.g. studies by Prohofsky {\em et al.}\cite{Pro95} or numerical
simulations by Lavery and collaborators\cite{Lave} and elasticity 
theories as the Worm-Like-Chain model\cite{Marko} and its recent
extensions\cite{Bou98}.

Our calculation gives access to the density of modes $\rho (\nu)$ and
some statistical properties of the normal modes as the dispersion
relation $\nu (q)$ and the autocorrelation length $1/\sigma (\nu )$. 
The dispersion relations provide the power spectrum of the modes which
exhibit a finite damping width $\sigma $ at ambient temperature. 

Let us summarize briefly our main quantitative result. Through a
rescaling of the density of modes taking into account the
light-to-vibrations coupling $C$, we have related the density of modes to
Raman intensity measurements. The choice of parameters $D=0.15 eV$,
$E=0.74 eV/{\AA}^2$ offers a good agreement with the experiments by Urabe
et al. (see figure~12.a and figure~2 in \cite{Ura81}). The range of frequencies
$50$~cm$^{-1} < \nu < 100$~cm$^{-1}$ corresponding to radial,
{\em i.e.} base pair stretching modes do not  qualitatively depend
on $C$. Furthermore, it remains roughly unchanged in the whole
interval of temperatures $0^o$C$<T<T_m$ where $T_m=77^o$C is the
denaturation temperature. Indeed, our model describes a homogeneous
sequence for which the melting transition is very abrupt and not
smooth as for disordered DNA. 

It would be very interesting to compare our theoretical results for
 $\sigma _r (\nu)$ and $\nu _r (q)$ with neutron scattering
experiments which to our best knowledge are not available for DNA in
solution over the range of frequencies mentioned above. A fundamental
feature of the modes is that the coherence length $\xi$ is of the
order of unity:  decorrelation between components of the same mode
takes place on a few Angstr{\"o}ms. This prediction agrees well with
results for the static correlation distance $\zeta$ obtained from
statistical mechanics calculations as seen Section~V.B. 

As for angular modes, the predicted characteristic frequencies
$\nu<25$~cm$^{-1}$ coincide with the Raman measurements
$\nu<16$~cm$^{-1}$\cite{Ura91}. The value of the coherence length in
the center of the spectrum, $\xi=1/\sigma \simeq 6.8 {\AA}$ is
compatible with data obtained through neutron scattering experiments
on DNA fibers \cite{Gri87}.  However, as far as angular modes are
concerned, the present approach suffers from two weaknesses . First,
we have not taken into account in the dynamical equations the viscous
forces that might become relevant at very low frequencies. Secondly,
INM can become unstable at small $\nu$ and the linearization
approximation we have used throughout the study reveals
dangerous. Further information about the non linear couplings between
modes would be extremely useful to circumvent this difficulty. It is
however not clear how such a study could be technically pursued.

\vskip .5cm
{\bf Acknowledgments~:} We are particularly grateful to W.~Baumruk and
P.Y.~Turpin for motivating and enligthening discussions on the
experimental aspects of this work. We also
thank J.F.~Leger, L.~Bourdieu, A.~Colosimo, D.~Chatenay, M.~Peyrard
and H.~Urabe for advises and discussions.

\appendix

\section{Variational calculation and self-consistent equation
for $Q$}

In this Appendix, we compute the Rayleigh-Ritz functional for both
the ground state (\ref{trwf}) and the first excited (\ref{trwf1}) (with
$\ell =1$) wave functions. We then derive the saddle-points equations
on the variance $Q(r)$ for both instantaneous modes families.

\subsection{Calculation of the ground state}

Inserting the Gaussian Ansatz (\ref{trwf}) in (\ref{rr2}), we obtain 
\begin{eqnarray}
{\cal R} [ \Psi _0 ^g ] &=& \frac{
\int _{r_{min}} ^\infty dr dr' \int _{0} ^\pi d\theta \;
T(r,r',\theta) \;\psi _0 (r)\; \psi _0 (r') \; \big\{
{\cal W} (r,r',\theta )  \big\} ^{-n/2} }
{ \int _{r_{min}} ^\infty dr\; \psi _0 (r) ^2 \;\big\{ 
-8\pi \; i\;  Q(r) \big\} ^{-n/2} } \nonumber \\
&=& \lambda _0 + \frac n2 \; r^g[Q] + O(n^2)
\label{rr50}
\end{eqnarray}
where ${\cal W}$ (\ref{opt16}) is the determinant of 
the two by two matrix ${\cal M}$ defined by
\begin{eqnarray}
{\cal M} = 
\left( \begin{array}{c c} Q(r) + \lambda + i\epsilon -
 d_0(r,r',\theta ) & d_1(r,r',\theta ) \\ d_1(r,r',\theta )   
&  Q(r') + \lambda + i\epsilon - d_0(r',r,-\theta )\end{array} \right)
\qquad ,
\label{matricem}
\end{eqnarray}
and 
\begin{eqnarray} 
r^g[Q] = && \lambda _0 \; \int _{r_{min}} ^\infty dr\;
\psi _0 (r) ^2 \ln Q(r) \nonumber \\
&&  - \int _{r_{min}} 
^\infty dr dr' \int _{0} ^\pi d\theta \; T(r,r',\theta)   
\;  \psi _0 (r)\; \psi _0 (r') \; \ln {\cal W} (r,r',\theta )
\label{rr51}
\end{eqnarray}
up to an irrelevant additional constant.  The vanishing condition
(\ref{rr4}) on the functional derivative of $r^g [Q]$ with respect to
$Q(r)$ leads to equation (\ref{opt1}).

\subsection{Calculation of the first excited state}

We now compute $\Lambda _1 ^g$ using Ansatz (\ref{trwf1}) and
expression (\ref{rr2}). The denominator of ${\cal R}[ \Psi ^g _1]$ reads
\begin{eqnarray}
\langle  \Psi ^g _1 |  \Psi ^g _1 \rangle &=& \int _{r_{min}} ^\infty dr
\int _{-\infty} ^\infty d\varphi \int d\vec x \; \psi _0 (r) ^2
\; x_1 ^2 \; \exp \left( \frac i2 Q(r) \; \vec x ^2 \right) 
\nonumber \\
&=& i \; C \;\int _{r_{min}} ^\infty dr \; \psi _0 (r) ^2 \; Q(r) ^{-1}  
\qquad ,
\end{eqnarray}
when $n\to 0$. $C= \int _{-\infty} ^\infty d\varphi$ could be
made finite by limiting the range of the twist angle. Such a
regularization is however
not necessary since $C$ also appears in the numerator of (\ref{rr2}),
\begin{eqnarray}
\langle  \Psi ^g _1 |{\cal T}|  \Psi ^g _1 \rangle &=& 
\int _{r_{min}} ^\infty dr dr' 
\int _{-\infty} ^\infty d\varphi d\varphi '  
\; T(r,r',\varphi -\varphi ') \;  \psi _0 (r)\; \psi _0 (r') \; \times
\nonumber \\
&& \qquad \qquad \times
\int d\vec x d\vec x' \; x_1 x_1 ' \; \exp \left[ \frac i4 
\left( \begin{array}{c} \vec x \\ \vec x ' \end{array} \right) ^\dagger
. {\cal M} .
\left( \begin{array}{c} \vec x \\ \vec x ' \end{array} \right)
\right]
\nonumber \\
&=& 2\, i \;C\; \int _{r_{min}} ^\infty dr dr' 
\int _0 ^\pi d\theta  \; T(r,r',\theta ) \;  \psi _0 (r)\; \psi _0 (r') 
\; \bigg( {\cal M} ^{-1} \bigg) _{12} (r,r', \theta ) \qquad ,
\end{eqnarray}
as the number of replicas $n$ vanishes. Using $\Lambda ^g _0 \to \lambda
_0$ when $n\to 0$, we obtain equation (\ref{lam1}).

\subsection{Case of base pairs vibrations}

For base pair vibrations, the second derivatives $d_0$ and $d_1$ are
given in (\ref{d0d1radius}) and do not depend on the relative twist
$\theta$ between successive base pairs. Inserting (\ref{d0d1radius})
into the extremization equation (\ref{opt1}) for $Q(r)$ and using the
definition (\ref{autremu}) of the effective radial transfer matrix 
$T_0$, we obtain
\begin{equation}
\frac {\lambda _0 \;\psi _0 (r)}{2\; Q(r)} = \int _{r_{min}} ^\infty
 dr'\; T_0 (r,r')\; \psi _0 (r') \;
\frac{ Q(r') + \lambda + i\epsilon - \frac 12 V''_m (r') -2 E} {{\cal
 W} _r (r,r')} \quad ,
\label{opt2}
\end{equation}
where
\begin{equation}
{\cal W} _r  (r,r') = \left(
Q(r) + \lambda + i\epsilon - \frac 12 V''_m (r) -2 E \right)
\left(
Q(r') + \lambda + i\epsilon - \frac 12 V''_m (r') -2 E \right) - 4 E^2 \ .
\label{opt22}
\end{equation}
The ratio (\ref{lam1}) of the first two eigenvalues of ${\cal T}$ is given 
by
\begin{equation}
\frac{\Lambda _1 ^g}{\Lambda _0 ^g} = - 4 E  \left\{
\int _{r_{min}} ^\infty dr dr' \; \frac{T_0 (r,r')}{{\cal W} _r
(r,r')} \;\psi _0 (r)\; \psi _0
(r') \right\}  \big/  \left\{ 
{\lambda _0 \int _{r_{min}} ^\infty dr\; \psi _0 (r) ^2 Q(r)^{-1} } \right\}
 \quad .\label{lam2}
\end{equation}

\subsection{Case of twist angle fluctuations}

For twist fluctuations, the elements $d_0,d_1$ of the matrix ${\cal D}
^{\cal C} _{\varphi}$ of second derivatives depend on radii $r,r'$ as
well as on the angle $\theta$. 
No simplification arises as in the base pairs case. The
saddle-point equation for $Q(r)$ is precisely equation (\ref{opt1})
but care must be paid to the backbone potential $V_b (r,r',\theta )$.
As can be seen from definition (\ref{vb}), the angular integral in
(\ref{opt1}) is restricted to twist angles $\theta$ fulfilling the
condition
\begin{equation} \label{crit}
\cos \theta > \frac{r^2 + r'^2 - L^2}{2\; r \; r'} \qquad .
\end{equation}
Indeed, when the inequality in (\ref{crit}) is not satisfied, the
potential $V_b$ is infinite, see discussion of Section II.A. 
The same prescription holds for the angular integral in (\ref{lam1}).

\newpage 
\begin{center}
{\large TABLES}
\end{center}
\vskip 2cm
$$
\begin{array}{||c|c|c||}
\hline \hline D  & E  & \Delta G  \\ (eV) & (eV/{\AA}^2) & (k_B T) \\ 
\hline \hline 0.15 & 0.74 & 0.762
 \\ \hline 0.16 & 4 & 0.825 \\ \hline 
0.17 & 12 & 0.691 \\  \hline \hline 
\end{array}\label{tavb}
$$
\vskip 1cm
{\bf Table 1:}
Three choices of the depths of the Morse potential $D$ and of
 the stacking stiffnesses $E$  giving the desired melting
 temperature $T_m=350$~K. The corresponding denaturation free-energies
$\Delta G$ are expressed in unit of $k_B T$ at $T=300$~K.


\newpage
\begin{center}
{\large FIGURE CAPTIONS}
\end{center}
\vskip 1cm
{\bf Figure 1:} The helicoidal DNA model: each base pair
is modelized through its radius $r_j$ and angle $\varphi_j$.
The axial distance $h_j$ between successive base pairs planes varies while 
the backbone length along the strands is fixed to $L$.
\vskip .5cm \noindent
{\bf Figure 2:} Ground state wave functions $\psi _0 (r)$ at $T=27^o$C 
for two choices of
the energetic parameters: $D=0.15 eV$, $E=0.74 eV/{\AA}^2$ (full curve) and
$D=0.16 eV$, $E=4 eV/{\AA}^2$ (dotted curve).
\vskip .5cm \noindent
{\bf Figure 3:} Dispersion relations at zero temperature. From bottom to top:
$\nu_{\varphi}(q)$ for torsion modes, $\nu_{r}(q)$ for radial modes with
two different choices of the parameters 
$D=0.15 eV$, $E=0.74 eV/{\AA}^2$ (full curve) and
$D=0.16 eV$, $E=4 eV/{\AA}^2$ (dotted curve).
\vskip .5cm \noindent
{\bf Figure 4:} Density of states at zero temperature.  
From left to right: torsion modes spectrum
$\rho _\varphi (\nu)$, radial modes spectra $\rho _r (\nu )$ for
$D=0.15 eV$, $E=0.74 eV/{\AA}^2$ (full curve) and $D=0.16 eV$, $E=4
eV/{\AA}^2$ (dotted curve).
\vskip .5cm \noindent
{\bf Figure 5:} Spectra $\rho _r (\nu )$ for radial modes at temperature
$T=27^o$C. Parameters are: $D=0.15 eV$, $E=0.74 eV/{\AA}^2$ (full curve)
and $D=0.16 eV$, $E=4 eV/{\AA}^2$ (dotted curve). Each spectrum is
normalized to one half. Note the small oscillations on the tails
of the curves due to numerical artifacts, see Section V.A.
\vskip .5cm \noindent
{\bf Figure 6:} Dispersion relations $\nu _r (q )$ for radial modes at
temperature $T=27^o$C.  Parameters are: $D=0.15 eV$, $E=0.74 eV/{\AA}^2$
(full curve) and $D=0.16 eV$, $E=4 eV/{\AA}^2$ (dotted curve).  For
comparison, dispersion relations at zero temperature are recalled
($E=0.74 eV/{\AA}^2$: dashed curve, $E=4 eV/{\AA}^2$: dash-dotted curve).
\vskip .5cm \noindent
{\bf Figure 7:} Inverse coherence length $\sigma _r (\nu )$ for radial
modes at temperature $T=27^o$C.  Parameters are: $D=0.15 eV$, $E=0.74
eV/{\AA}^2$ (full curve) and $D=0.16 eV$, $E=4 eV/{\AA}^2$ (dotted
curve).
\vskip .5cm \noindent
{\bf Figure 8:} Spectra $\rho _\varphi (\nu )$ for angular modes for $D=0.16
eV$, $E=4 eV/{\AA}^2$. The curve for $D=0.15 eV$, $E=0.74 eV/{\AA}^2$ is
indistinguishable from the latter and from the approximation
(\ref{dt0},\ref{dt1},\ref{opt10}). The spectrum is normalized to one
half. Negative frequencies represent unstable modes according to the
convention exposed in Section III.B.
\vskip .5cm \noindent
{\bf Figure 9:} Dispersion relation $\nu _\varphi (q )$ for angular modes
(full curve). For comparison, the dispersion relation at zero
temperature is recalled (dashed curve). 
Negative frequencies represent unstable modes
according to the convention exposed in Section III.B. Note the
inflection point at $\nu =0$ due to this representation, see Section~V.C.
\vskip .5cm \noindent
{\bf Figure 10:} Inverse coherence length $\sigma _\varphi (\nu )$ for angular
modes. Negative frequencies represent unstable modes
according to the convention exposed in Section III.B. Note the
inflection point at $\nu =0$ due to this representation, see Section~V.C.
\vskip .5cm \noindent
{\bf Figure 11:} Spectrum $\rho (\nu )$ for both radial and angular modes 
 Parameters are: $D=0.15 eV$, $E=0.74 eV/{\AA}^2$ 
(full curve) and $D=0.16 eV$, $E=4 eV{\AA}^2$ (dotted curve). 
Each spectrum is normalized to unity
\vskip .5cm \noindent
{\bf Figure 12:} Rescaled spectra for different 
 light-to-vibrations coupling $C$ for parameters $D=0.15 eV$, $E=0.74
 eV/{\AA}^2$ ({\bf a}) and $D=0.16 eV$, $E=4 eV/{\AA}^2$ ({\bf b}). 
Original spectra $\rho (\nu)$ are recalled (full lines) and 
rescaling functions are: $C(\nu)=1$ (dotted curve), $C(\nu)=\nu$ 
(dashed curve) and $C(\nu)= \nu ^2$ (dashed-dotted curve). Vertical 
units are arbitrary, all curves have been multiplied by a constant to
meet (and equal unity) in $\nu =50$~cm$^{-1}$.

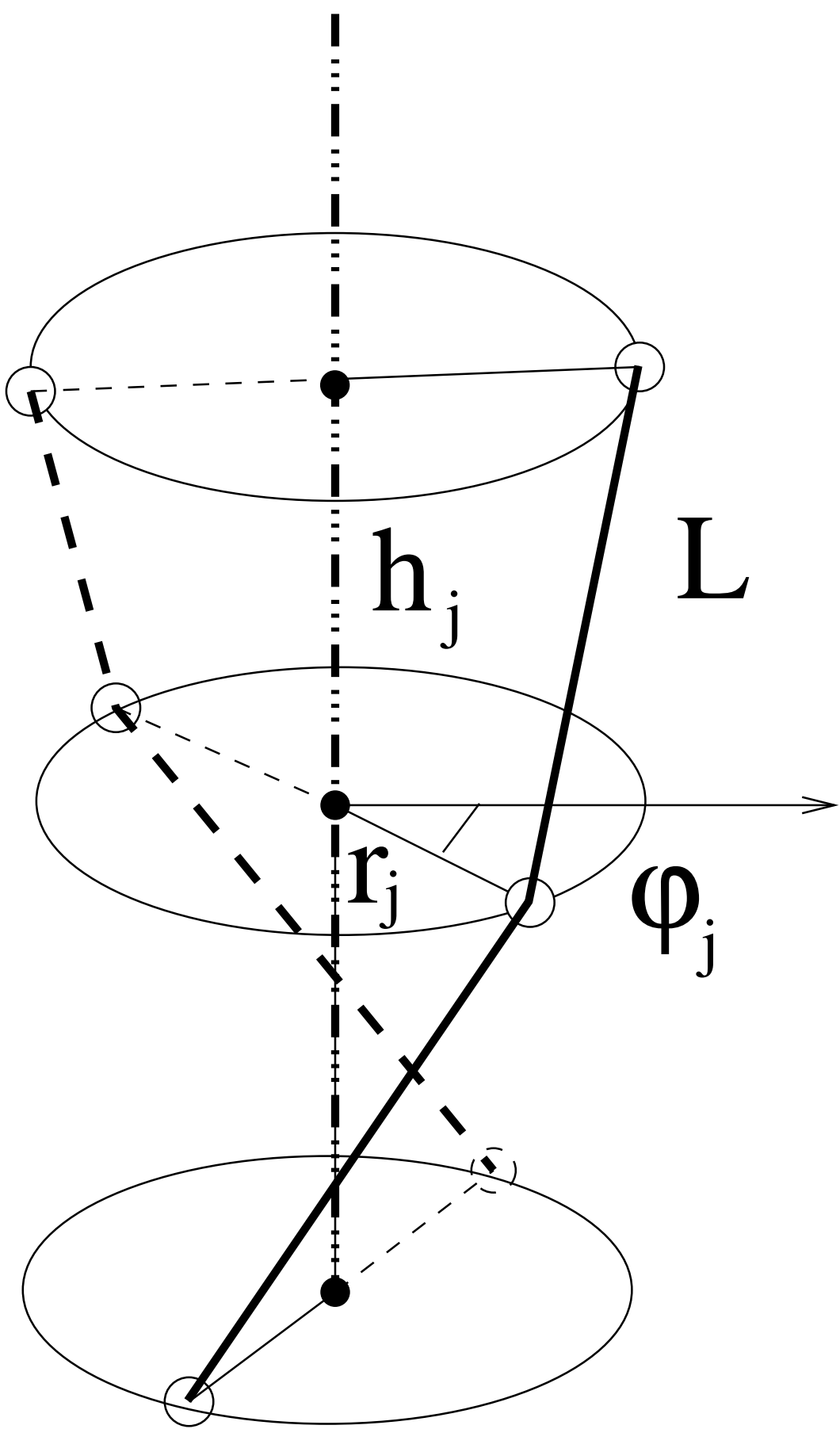
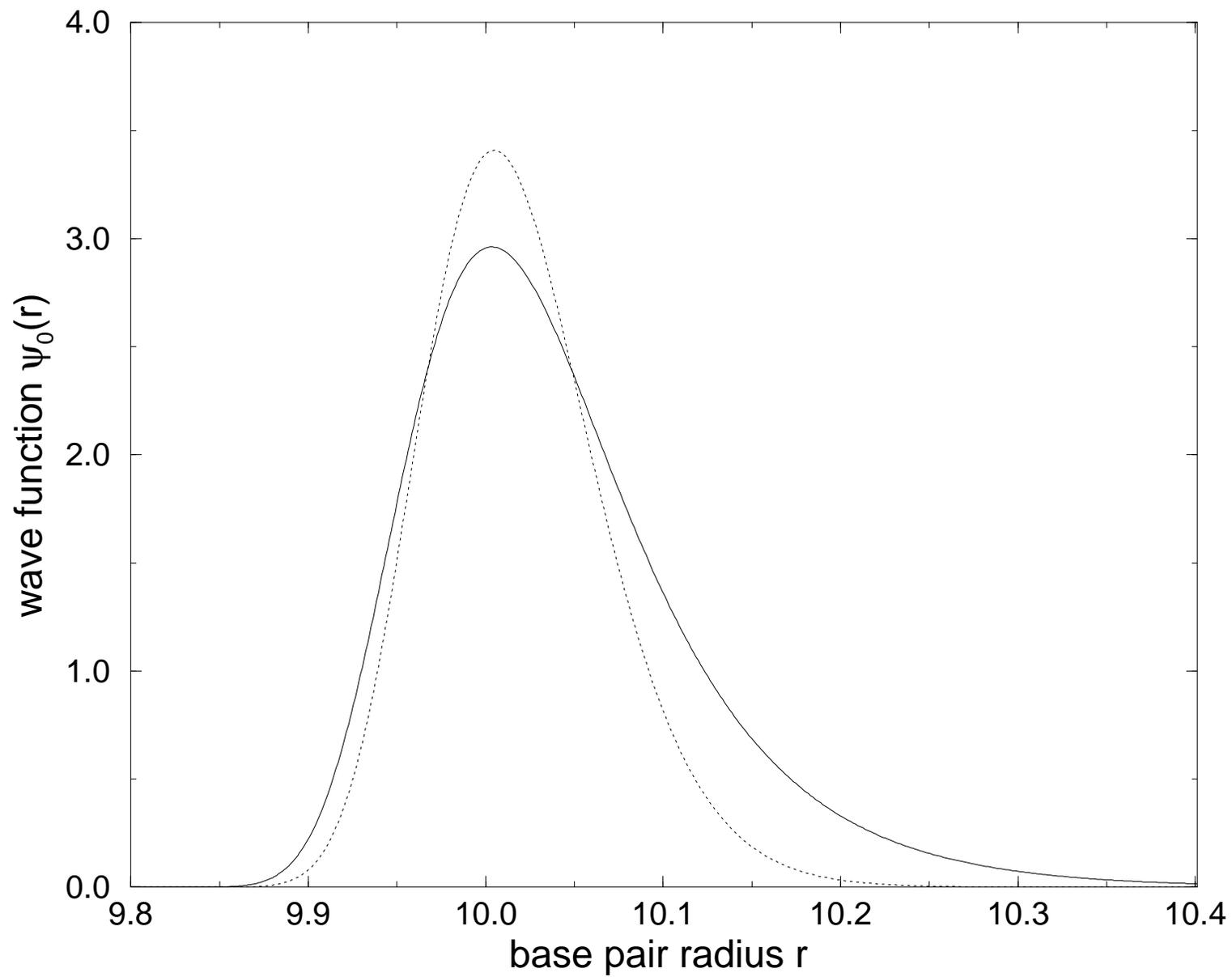
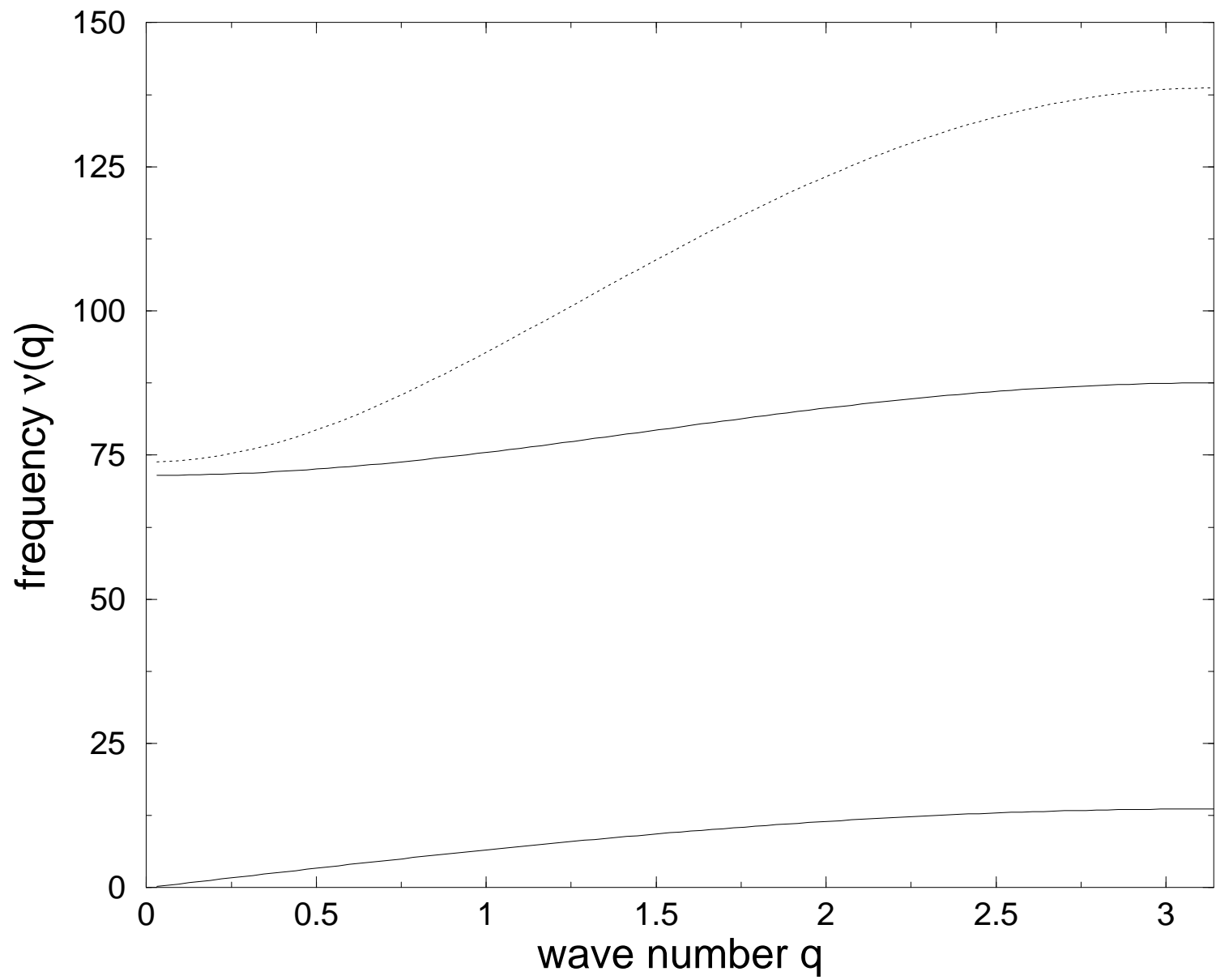
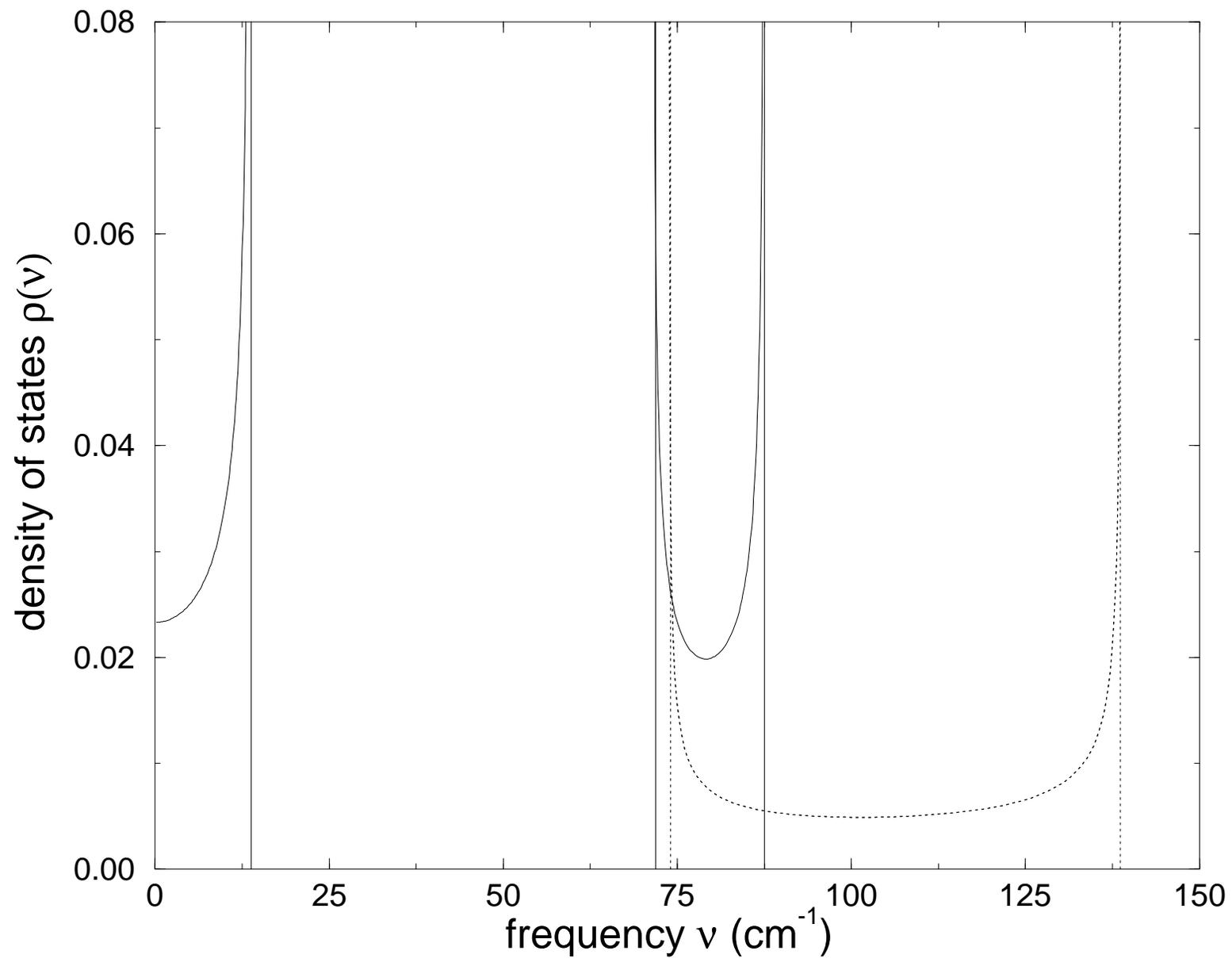
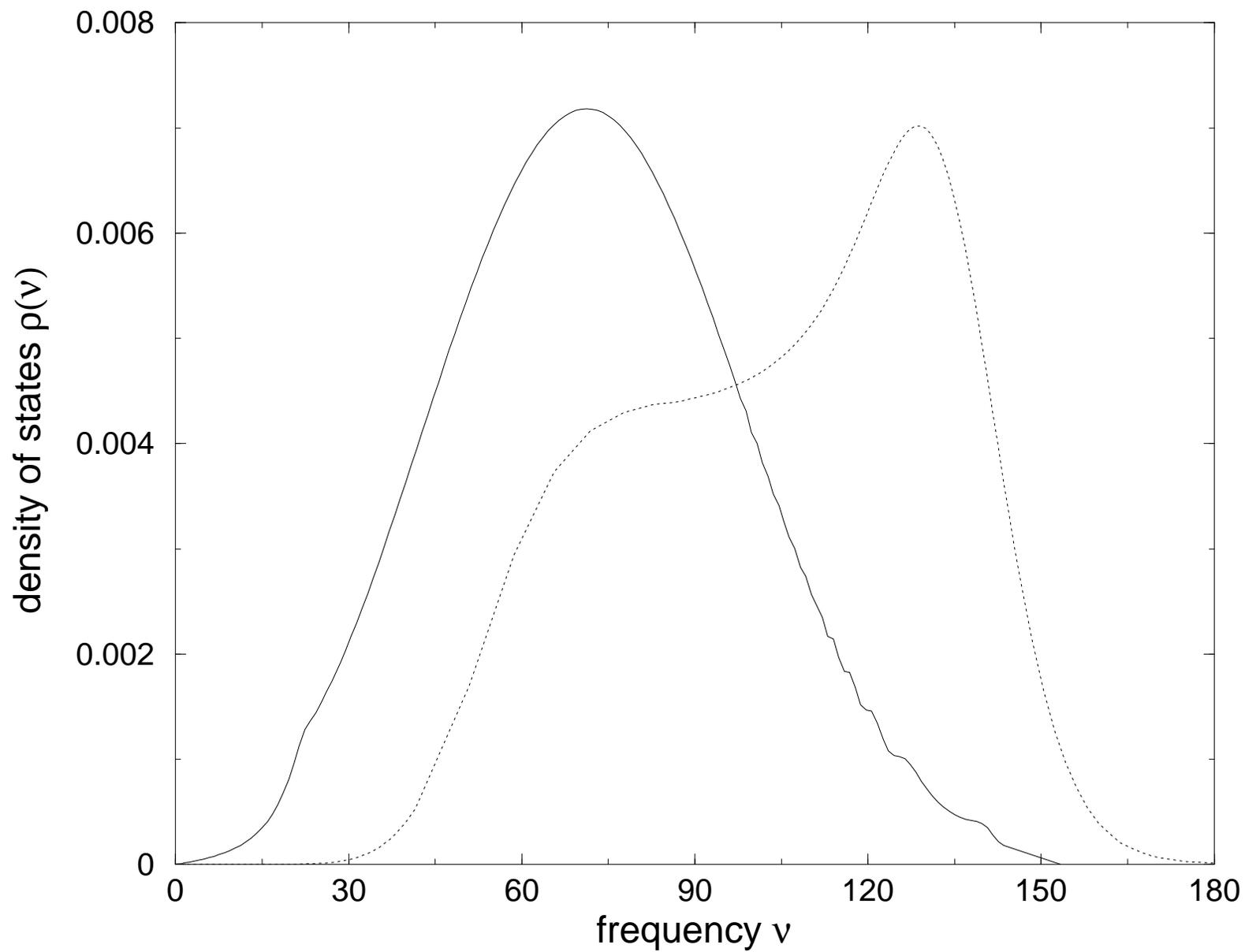
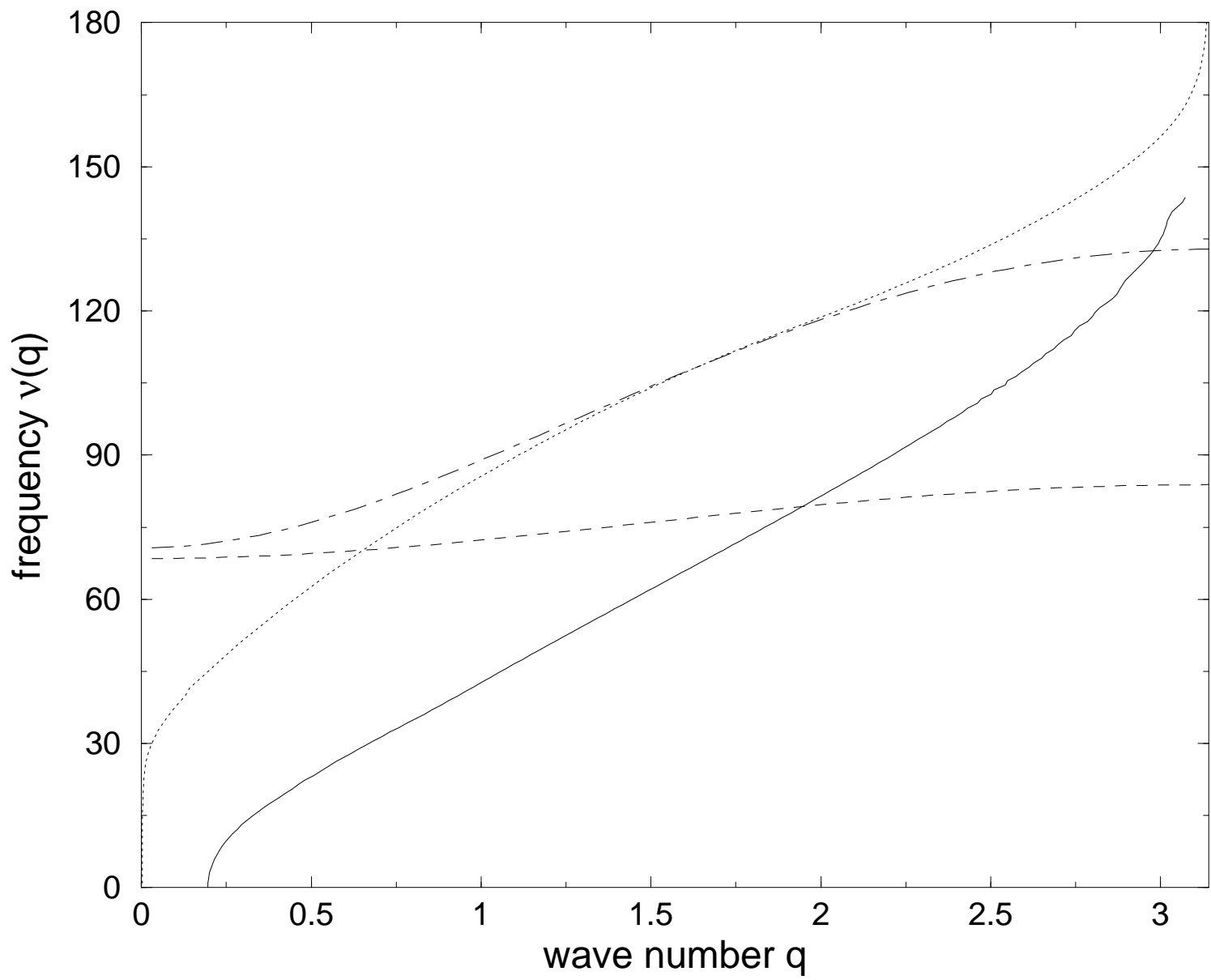
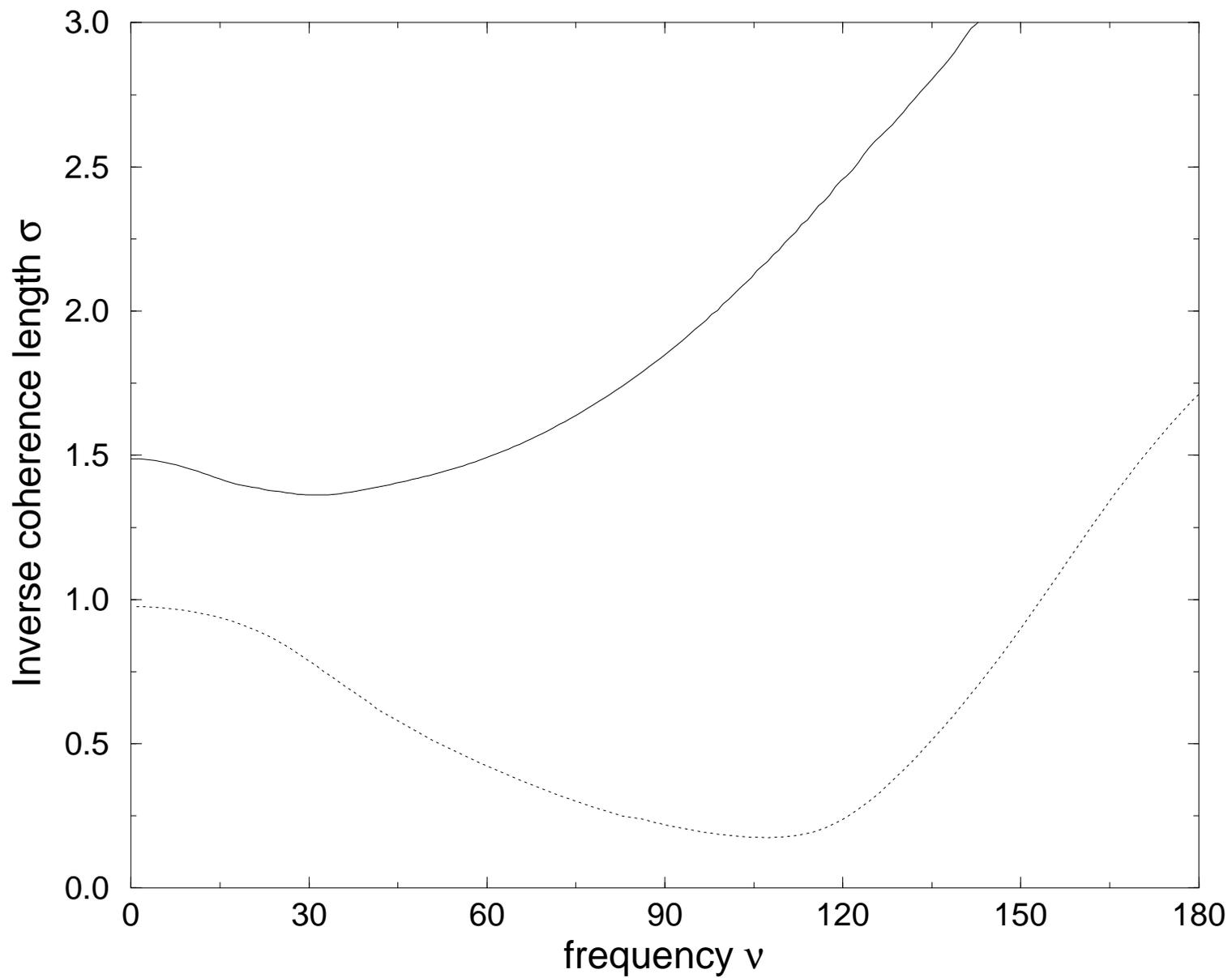
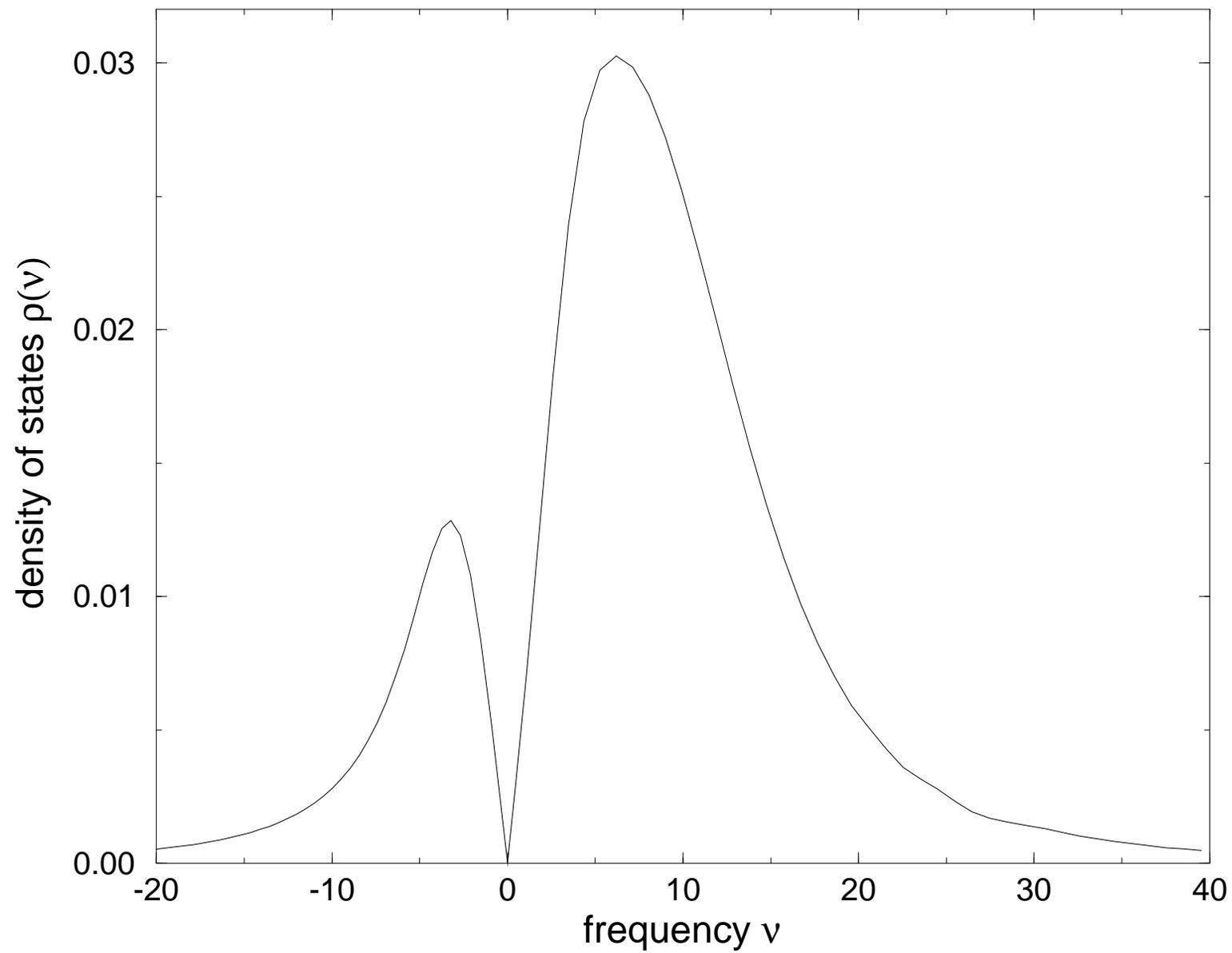
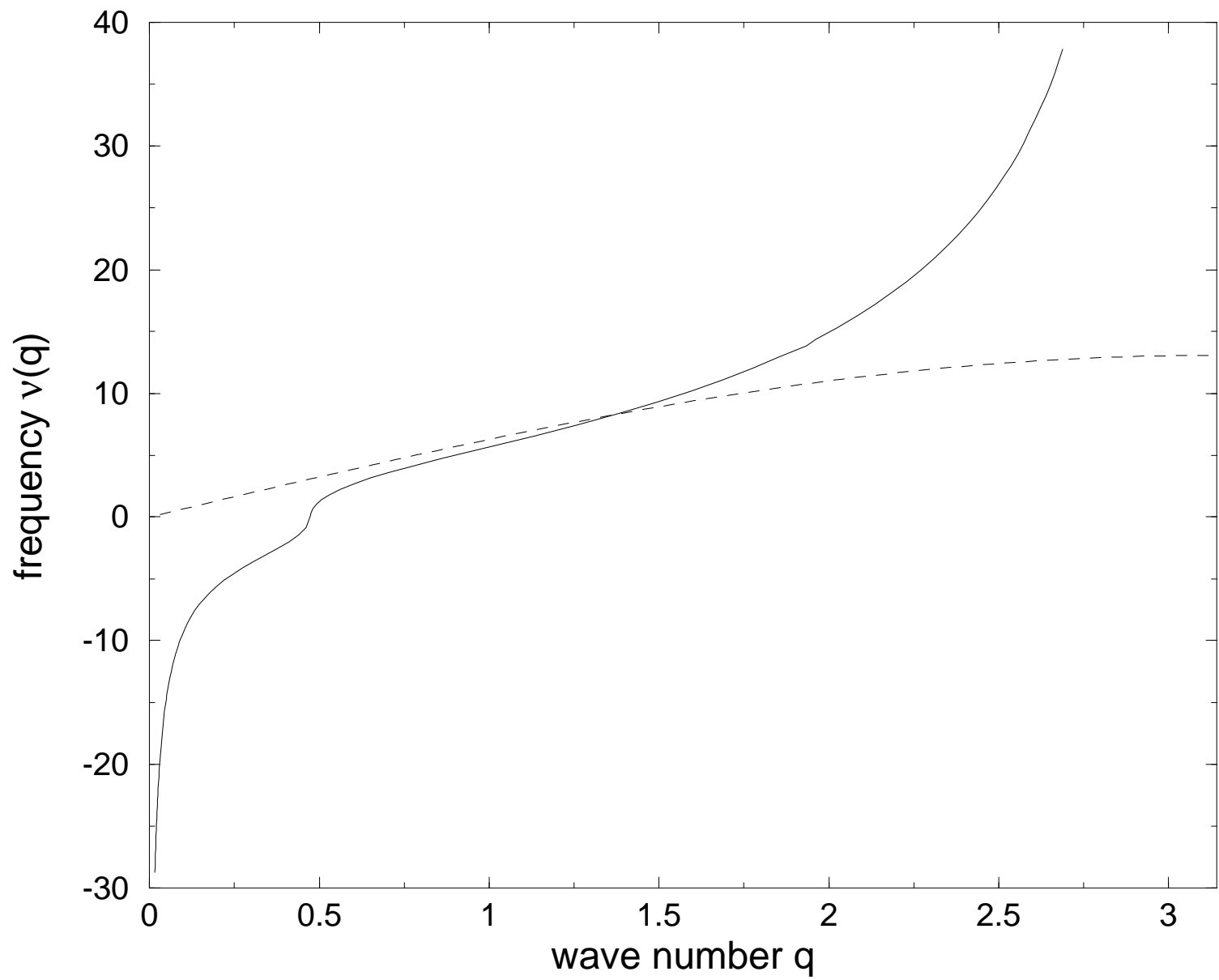
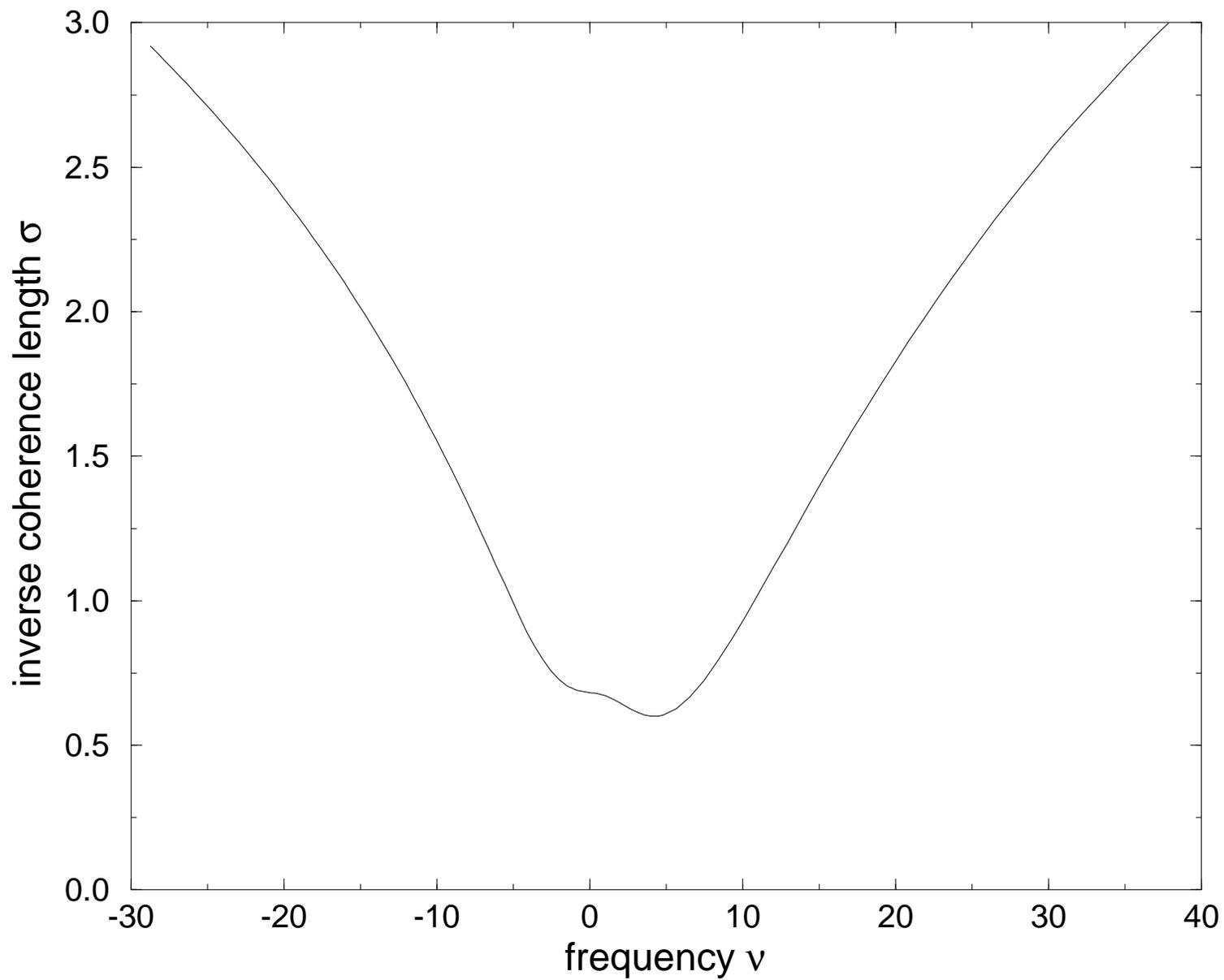
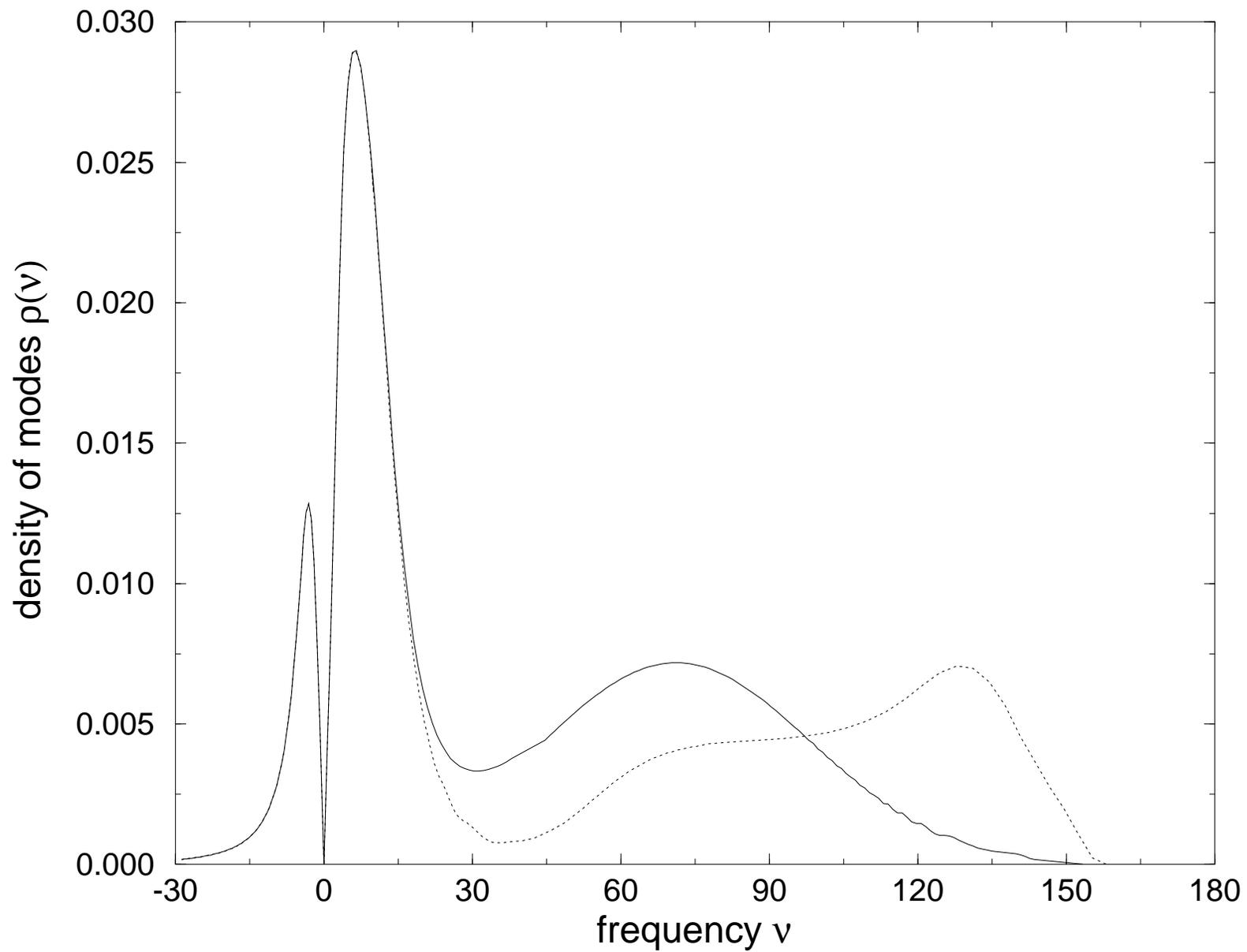
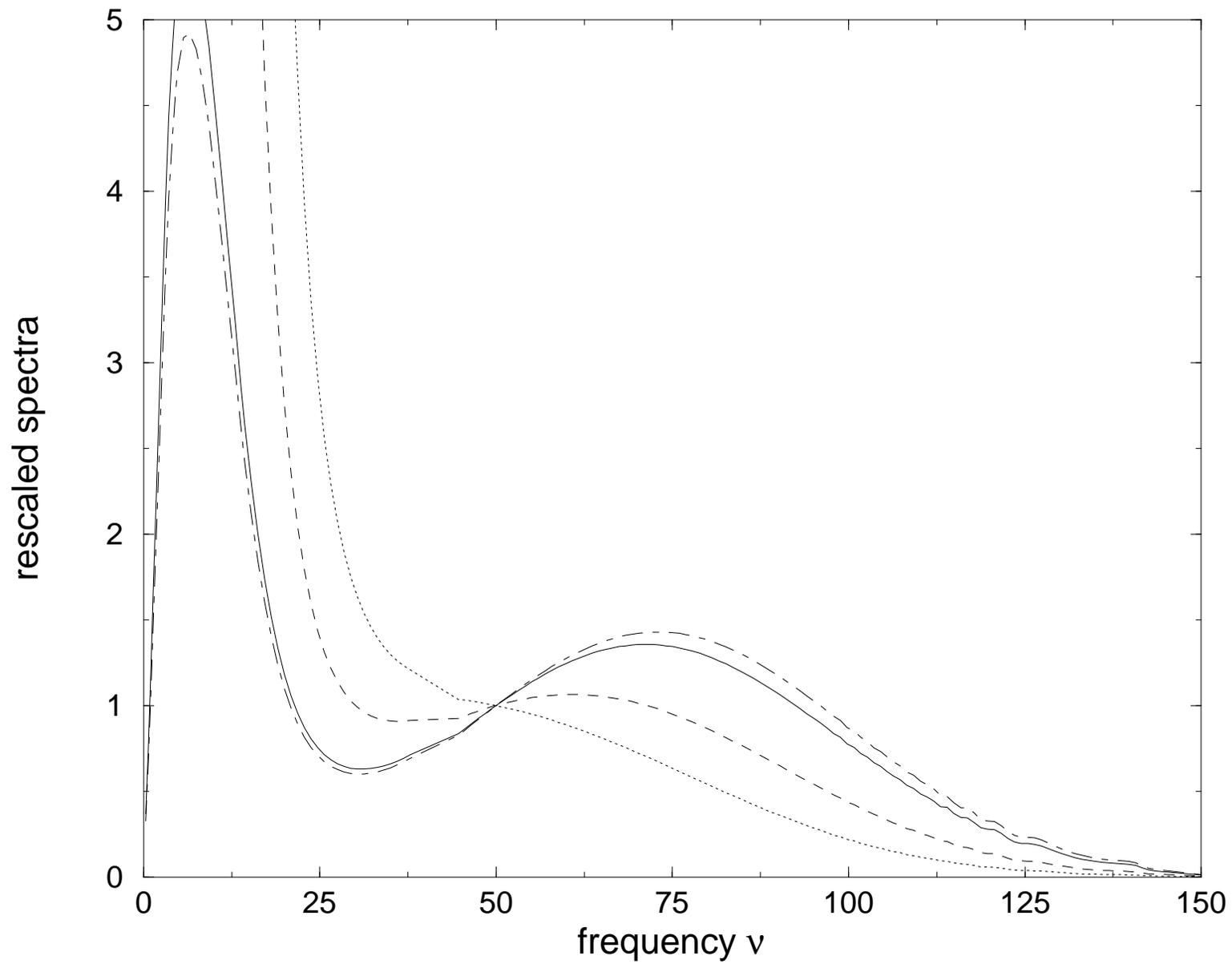
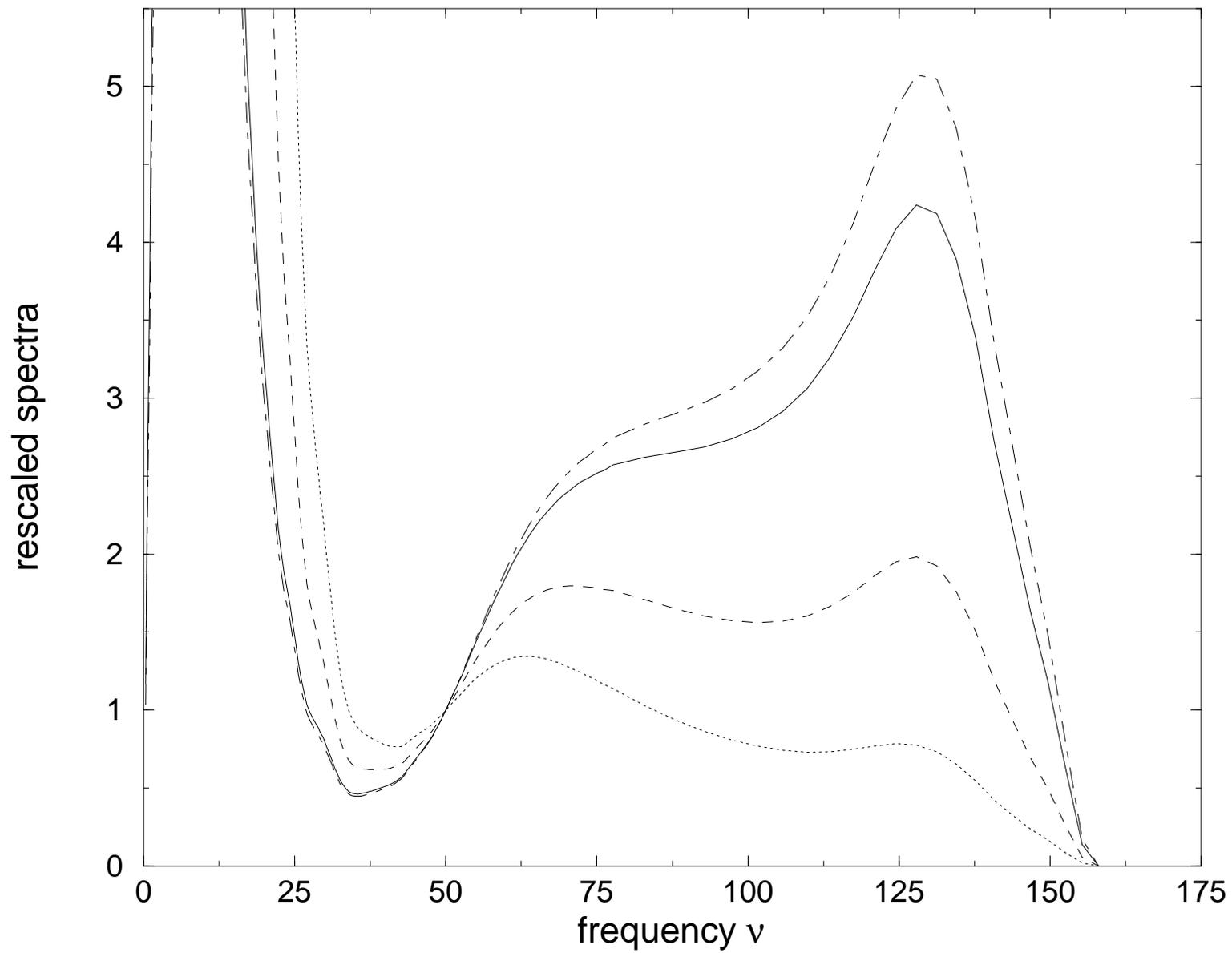
\end{document}